\documentclass[aps,pre,floats,floatfix,twocolumn]{revtex4}

\usepackage{ifthen}
\usepackage{ifpdf}
\usepackage{color}

\ifpdf
\usepackage{graphicx}
\usepackage{epstopdf}
\else
\usepackage{graphicx}
\usepackage{epsfig}
\fi

\usepackage{latexsym}
\usepackage{amsmath}
\usepackage{amssymb}
\usepackage{bm}
\usepackage{wasysym}

\newcommand{\eexp}{\mbox{e}^}
\newcommand{\bra}{\left\langle}
\newcommand{\ket}{\right\rangle}

\newcommand{\mass}{\mathsf{m}} 
 
\newcommand{\hbarr}{\hbar} 

\newcommand{\tbox}[1]{\mbox{\tiny #1}}
 
\newcommand{\mylabel}[1]{\label{#1}} 
\newcommand{\be}[1]{\begin{eqnarray}\ifthenelse{#1=-1}{\nonumber}{\ifthenelse{#1=0}{}{\mylabel{e#1}}}}
\newcommand{\ee}{\end{eqnarray}}

\newcommand{\Eq}[1]{\textcolor{blue}{Eq.~\!\!(\ref{#1})}}
\newcommand{\Sec}[1]{\textcolor{blue}{Sec.~\!\!(\ref{#1})}}
\newcommand{\App}[1]{\textcolor{blue}{App.~\!\!(\ref{#1})}}
\newcommand{\Fig}[1]{\textcolor{blue}{Fig.~\!\!\ref{#1}}}  
\newcommand{\hide}[1]{} 
\newcommand{\mycite}[1]{\textcolor{blue}{\cite{#1}}}
\newcommand{\sect}[1]{\vspace{2mm} {\bf #1.-- }}

\begin{document}

\title{``Weak Quantum Chaos'' and its resistor network modeling}


\author{Alexander Stotland$^1$, Louis M. Pecora$^2$ and Doron Cohen$^1$}

\affiliation{
\mbox{$^1$Department of Physics, Ben-Gurion University, Beer-Sheva 84105, Israel}
\mbox{$^2$Code 6362, Naval Research Lab, Washington DC 20375, USA} 
}


\begin{abstract}
Weakly chaotic or weakly interacting systems have a wide regime 
where the common random matrix theory modeling does not apply. 
As an example we consider cold atoms in a nearly integrable optical 
billiard with displaceable wall (``piston"). The motion is completely 
chaotic but with small Lyapunov exponent. The Hamiltonian matrix  
does not look like one taken from a Gaussian ensemble, but rather 
it is very sparse and textured. This can be characterized by 
parameters~$s$ and~$g$ that reflect the percentage of large elements, 
and their connectivity, respectively. For~$g$ we use a resistor network 
calculation that has a direct relation to the semi-linear response 
characteristics of the system, 
hence leading to a novel prediction regarding the EAR 
of cold atoms in optical billiards with vibrating walls.
\end{abstract}

\maketitle


\section{Introduction}

So called "quantum chaos" is the study of quantized
chaotic systems. Assuming that the classical dynamics
is fully chaotic, as in the case of a billiard 
with convex walls (\Fig{f11}), one expects the Hamiltonian 
to be like a {\em random matrix} with elements that have a Gaussian
distribution. This is, of course, a sloppy statement,
since any Hamiltonian is diagonal in some basis. 
The more precise statement behind random matrix theory (RMT) 
is following \mycite{mario1,mario2,mario3,mario4,prosen1,prosen2}.
Assume that there is a Hamiltonian $\mathcal{H}$ 
that generates chaotic dynamics, 
and consider an observable $F$ that has some classical  
correlation function $C(t)$, with some correlation time $t_{\tbox{R}}$.
Then the matrix representation $F_{nm}$ in the basis
of $\mathcal{H}$ looks like a random banded matrix. 
The bandwidth is $\hbarr_{\tbox{Planck}}/t_{\tbox{R}}$. 
If $t_{\tbox{R}}$ is small, such that the bandwidth 
is large compared with the energy window of interest, 
then the matrix looks like taken from a Gaussian ensemble.

Our objective is to analyze the energy absorption rate (EAR) 
of billiards with vibrating walls, 
which is related to past studies of Nuclear friction
\mycite{wall1,wall2,wall3}.    
However, our interest is focused on 2D optical billiards 
\mycite{nir1,nir2,nir3,nir4,kbw} 
whose geometrical shape can be engineered. 
In this problem $\mathcal{H}$ is the Hamiltonian 
of the non-driven billiard, while $F_{nm}$ 
is the perturbation matrix due to the wall displacement.
If the driving is not too strong we expect a linear relation  
\be{1865}
\dot{E} \ \ = \ \ G \ \dot{f}^2
\ee
where $\dot{f}$ is the RMS value of the vibrating wall velocity. 
If one further assumes that the billiard is strongly chaotic, 
then $G=G_0$ can be determined (\Eq{e99}) from simple 
kinetic considerations as in \mycite{wall1,wall2,wall3},
leading to a variation of the so-called {\em Wall-formula}.
Note that there is a strict analogy here with the Drude 
formula and the Joule law.

We consider completely chaotic billiards \mycite{bunimovich,young},
with no mixed phase space, but we assume that they are only {\em weakly} chaotic 
\mycite{bouncing1,bouncing2,triang,spectral,Backer,wqc1,wqc2,cqb}. 
This means that $t_{\tbox{R}}$ is much larger than the ballistic time $t_{\tbox{L}}$. 
Consequently, the EAR coefficient is 
\be{0}
G \ \ = \ \ g \ G_0
\ee
with $g\ne1$. In the classical analysis $g=g_c$ is related 
to classical correlations between the collisions with the 
vibrating walls. In the quantum analysis the first tendency
is to assume  $g \approx g_c$. In contrast to that 
we would like to highlight the possibility to observe $g \ll g_c$.
This is the case if we have {\em weak quantum chaos} (WQC) circumstances, 
in which the traditional RMT modeling does not apply, 
meaning that $F_{nm}$ does not look like a typical 
random matrix. Rather, the distribution of its elements 
is log-wide (resembles a log-normal distribution), 
and it looks very sparse, as expected 
from Refs.\mycite{sparse1,sparse2,sparse3,sparse4}. 
Consequently, the analysis of the EAR
has to go beyond the familiar framework 
of {\em linear response theory} (LRT).

WQC circumstances are encountered in the analysis 
of any weakly chaotic or weakly interacting system. 
In the WQC regime the matrix $F_{nm}$ is formed of 
elements that have a log-wide distribution. 
The implied sparsity is important for the analysis 
of the EAR \mycite{kbw,cqb}, as expected 
from {\em semi-linear response theory} (SLRT) \mycite{kbr,slr,kbd}.
The main idea behind the theory is demonstrated 
in \Fig{f10}: one observes that the energy absorption 
process requires {\em connected} sequences of transitions 
between the energy levels of the system.
Accordingly, the calculation of EAR requires 
a semi-linear resistor network calculation.
   
We can characterize the sparsity of the perturbation matrix 
by parameters~${s}$ and~${g_s}$ that reflect the percentage 
of large elements, and their connectivity, respectively. 
The parameter~$g_s$ is defined through a resistor network calculation, 
and has a direct relation to the semi-linear response characteristics 
of the system, namely
\be{0}
g \ \ = \ \ g_s \ g_c
\ee
For a strictly uniform matrix ${g_s=s=1}$, 
for a Gaussian matrix ${s=1/3}$ and ${g_s\sim1}$, 
while for sparse matrix ${s,g_s\ll1}$. 
We would like to explore the dependence of $g_c$ and $g_s$ 
on the parameters $u$ and $h$ of the system. 
Disregarding the physical motivation, this exploration 
is mathematically interesting, because it introduces 
a ``resistor network" perspective into RMT studies. Hence, it is complementary 
to the traditional spectral and intensity statistics investigations.

\subsection{Generic parameters}

For a billiard of linear size~$L$ that has walls with  
radius of curvature~$R$, as in \Fig{f11}, 
the Lyapunov time and the ballistic time are 
\be{0}
t_{\tbox{R}}=& R/v_{\tbox{E}}  \ \ \ \ \ &\mbox{[Lyapunov time]} \\
t_{\tbox{L}}=& L/v_{\tbox{E}}  \ \ \ \ \ &\mbox{[Ballistic time]} 
\ee
where ${v_{\tbox{E}}=(2E/\mass)^{1/2}}$ is the velocity of the particle. 
The quantization introduces an additional length scale into the problem, 
the de~Broglie wavelength 
\be{0}
\lambda_{\tbox{E}} \ \equiv \ \frac{2\pi}{k_{\tbox{E}}} 
\ \equiv \ \frac{h_{\tbox{Planck}}}{\mass v_{\tbox{E}}}   
\ee
Accordingly, the minimal model for the purpose of our study  
is featured by two {\em small} dimensionless parameters:  
\be{0}
u = & L/R & = \mbox{[degree of deformation]} 
\\ \label{e500}
\hbarr =& \lambda_{\tbox{E}}/L &= \mbox{[scaled Planck]} \ = 2\pi/(k_{\tbox{E}}L)  
\ee
With the two classical time scales $t_{\tbox{L}}$ and $t_{\tbox{R}}$
one may associate two frequencies, while quantum mechanics adds an additional 
frequency that corresponds to the mean level spacing:
\be{0}
\Delta_{\tbox{L}} =& 2\pi/t_{\tbox{L}}  & \\
\Delta_{\tbox{R}} =& 2\pi/t_{\tbox{R}}  &= \ u\,\Delta_{\tbox{L}} \\
\Delta_{0} \equiv &  2\pi/t_{\tbox{H}}  &= \ (\hbarr/2\pi)^{d{-}1}\,\Delta_{\tbox{L}} 
\ee
where $d{=}2$ is the dimensionality of the billiard.  
The WQC circumstances that we would like to consider 
are characterized by the following separation of scales: 
\be{0}
\Delta_{0}, \ \Delta_{\tbox{R}} \ \ll \ \Delta_{\tbox{L}}
\ \ \ \ \ \ \ \ \ \mbox{[WQC]}
\ee
However, this is not a sufficient condition to observe WQC.
The identification of the WQC regime in the ${(u,h)}$ space 
is an issue that we would have to address.

\subsection{Detailed outline}

\sect{The model}
In \Sec{s2} we define the model and explain the numerical procedure.
Schematically the Hamiltonian of the system can be written as 
\be{1}
\mathcal{H}_{\tbox{total}}  &=& \mathcal{H} - f(t) F  \\ &=& \mathcal{H}_0 + U  - f(t) F
\ee
where $\mathcal{H}_0$ describes the undeformed rectangular box, 
and $U$ describes the deformation of the fixed walls, 
and $F$ is the perturbation due to the displacement~$f(t)$ 
of the moving wall (piston). 
The geometry of the billiard is characterized by~$u$, 
while $\hbarr$ defines via \Eq{e500} the energy~window of interest.

\sect{Eigenstates}
Given $u$ and $\hbarr$ we find the ordered eigenenergies $E_n$ 
of the Hamiltonian~$\mathcal{H}$, within the energy~window 
of interest. This is done using the boundary element method~\mycite{boundary}. 
A representative eigenstate is presented in \Fig{f12}.
{If the deformation is small} it is meaningful to 
represent it in the basis which is defined by $\mathcal{H}_0$.
See for example \Fig{f13}. As the deformation becomes larger 
more and more levels are mixed as demonstrated in \Fig{f14}.

\sect{Perturbation matrix}
Once we have the eigenstates we calculate the matrix elements~$F_{nm}$ 
of the perturbation term~$F$, within the energy window of interest. 
An image of a representative matrix is shown in \Fig{f31}.  
The distribution of its elements is definitely {\em not} Gaussian, 
as shown in \Fig{f32}. In fact we see that the statistics 
of~$|F_{nm}|^2$ resembles a log-normal distribution.

\sect{Bandprofile}
In order to characterize the bandprofile of the perturbation 
matrix we define in \Sec{s2} spectral functions $\bar{C}_a(r)$ 
and $\bar{C}_s(r)$ that are displayed in \Fig{f33}. 
We explain how $\bar{C}_a$ is semiclassically 
related to the power spectrum $\tilde{C}(\omega)$ of the collisions, 
and how $\bar{C}_s$ gives an indication for the 
sparsity of the matrix.

\sect{Sparsity}
We characterize the sparsity and the texture 
of the matrix by a parameter $g_s$ 
that is defined in \Sec{s3}. 
The numerical results for $g_s$ 
are presented in \Fig{f34}. 
This characterization cares about the connectivity 
of the matrix elements, and is based on a calculation 
of a resistor network average. 
Some further details with respect to the 
resistor network average are given in \App{a0}.

\sect{Classical analysis}
In \Sec{s4} we provide a detailed analysis 
of the classical power spectrum $\tilde{C}(\omega)$.
In particular, we derive an expression for 
the zero frequency peak, which reflects  
the long time correlations of the bouncing 
trajectories in our weakly chaotic billiard. 
\App{a1} provides optional perspectives 
with regard to this classical calculation.

\sect{Quantum analysis}
In \Sec{s5} we use 1st order perturbation theory 
for the analysis of the eigenstates of the deformed 
billiard, and hence get an approximation for $\tilde{C}(\omega)$. 
The validity of this approximation is very limited.
We therefore fuse perturbation theory with 
semiclassical considerations in \Sec{s6}.  
This allows to obtain some practical approximations 
for $g_c$ and $g_s$.

\sect{The WQC regime}
Eventually we turn to define the borders of the WQC regime 
in the ${(u,h)}$ parameter space.
This is also an opportunity to make a connection with 
previous works that concern spectral and intensity 
statistics for such type of weakly chaotic billiards.

\sect{Implications}
The relevance of the resistor network analysis
to the EAR calculation is clarified in \Sec{s8}, 
and the experimental feasibility of observing 
the implied SLRT anomaly is discussed in \Sec{s9}.  
A broader perspective with respect to EAR predictions 
is presented in \Sec{s10}. In particular, we 
clarify how to bridge between what looks like 
contradicting results with regard to diffusive
and diffractive systems, as opposed to ballistic billiards.

\section{The model, Numerics}

\label{s2}

The Hamiltonian of the system is 
\be{0}
\mathcal{H}_{\tbox{total}} 
= \frac{\bm p^2}{2\mass} + W_{\tbox{box}}\big(x,y\big) + W_f\big(x {-} f(t)D_f(y)\big) 
\ee
We write it formally as in \Eq{e1},  
where $\mathcal{H}_0$ describes the undeformed rectangular box ${L_x \times L_y}$.
The ballistic time is defined as $t_{\tbox{L}}=L_x/v_{\tbox{E}}$, 
which is the typical time for successive collisions with the piston.    
The term $U$ in the Hamiltonian is due to a deformation $D_u(y)$ 
of the left static wall. The amplitude of this deformation 
is $D_0 \sim L_y^2/R$, while ${u \equiv L_y/R}$ is conveniently  
defined as the dimensionless deformation parameter. 
The driving term $-f(t)F$ is due 
to the deformation $f(t)D_f(y)$ of the right wall, 
leading to the identification 
\be{0}
F \ \ = \ \ D_f(y) W_f'(x)
\ee
For parallel displacement of a ``piston" we set $D_f(y){=}1$.  
Note that $f(t)D_f$ unlike $D_u$ is assumed to be small 
compared with the de~Broglie wavelength. 

For a billiard system with ``hard" walls the potential 
is zero inside the box, and become very large 
outside of the box. Accordingly, it is assumed 
that $W_f(r)=0$ for $r<0$, 
with a steep rise as $r$ becomes $>0$.  
Accordingly, the penetration distance upon collision 
is much smaller compared with the linear dimension~$L$ of the box. 
Say that the force which is exerted on the particle 
by the piston is $W_f'(r)=F_0$ for $r>0$ and zero otherwise,
then it is assumed that $E/F_0 \ll L$, where~$E$ 
is the kinetic energy of the particle inside the box. 
Below we take the limit $F_0\rightarrow\infty$. 

Following \mycite{frc,dil,wlf} we discuss 
the definition and the calculation of the 
spectral function $\tilde{C}(\omega)$
that describes the fluctuations of~$F$ 
in the non-driven billiard system. We first discuss 
the classical and then turn to the quantum.

\sect{Classical}
In the classical context the Hamiltonian $\mathcal{H}$ can be used to 
generate a trajectory ${(t_j,y_j,\theta_j)}$, 
where $j$ labels successive collisions with the piston 
at ${(0,y_j)}$ with incident angle $\theta_j$. 
Consequently, the associated~$F(t)$ consists of
impulses of height $F_0$ whose duration 
is $2\mass v_{\tbox{E}} \cos(\theta_j)/F_0$. 
In the hard wall limit one can write formally 
\be{6}
F(t) \ \ = \ \  \sum_j 2\mass v_{\tbox{E}} \ \cos(\theta_j) \ D_f(y_j) \
\delta(t-t_j) 
\ee
Assuming ergodic motion, the auto-correlation function of $F(t)$ 
can be calculated from the time dependence of a single 
trajectory that has some long duration $t_{\infty}$  
\be{0}
C(t) = \langle F(t) F(0) \rangle = \overline{F(t)F(0)}
\ee
The associated power spectrum is:
\be{11}
\tilde{C}_{\tbox{cl}}(\omega)  
&\equiv& 
\int_{-\infty}^{\infty} C(t) \,\eexp{i\omega t} dt
=
\frac{1}{t_{\infty}} |\tilde{F}(\omega)|^2 
\\
&=& 
\frac{1}{t_{\infty}}  
\left| 
\sum_j 2mv_{\tbox{E}} \cos(\theta_j)  D_f(y_j) \eexp{i\omega t_j}
\right|^2
\ee
If we regard the impulses as uncorrelated we get the result
\be{13}
\tilde{C}_{\tbox{cl}}(\omega)  
\ \  \rightarrow \ \ 
\left[\frac{8}{3\pi} \frac{\mass^2 v_{\tbox{E}}^3}{L_x}\right] 
\ \ \equiv \ \ C_{\infty} 
\ee
which holds in the $\omega\rightarrow\infty$ limit.
More details about this calculation and its  refinement 
will be presented in \Sec{s4}.

\sect{Quantum}
The unperturbed energy levels of the rectangular box are 
\be{0}
E_{\bar{\bm{n}}} \ = \ E_{n_xn_y} \ = \ \frac{1}{2\mass}\left[\left(\frac{\pi n_x}{L_x}\right)^2+\left(\frac{\pi n_y}{L_y}\right)^2\right]
\ee
with the mean level spacing 
\be{0}
\Delta_0 \ \ = \ \ \frac{2\pi}{\mass L_xL_y}
\ee
For a given deformation we diagonalize~$\mathcal{H}$,
and find the ordered eigenenergies~$E_n$ with ${n=1,2,3,...}$, 
within an {\em energy~window} of interest which 
is characterized by the dimensionless parameter~$\hbarr$.
This is done using the boundary element method~\mycite{boundary}. 
Each eigenstates $\psi(x,y)$ is represented by 
a boundary function $\varphi(y) \equiv \psi'(0,y)$,
where the normal derivative is with respect to~$x$ 
at the position ${x{=}0}$ of the piston.  
Consequently, the matrix elements of~$F$ are 
\be{16}
F_{nm} = 
-\frac{1}{2\mass}\int_0^{L_y} \varphi^{(n)}(y)\varphi^{(m)}(y) \ D_f(y) dy
\ee
Given $F_{nm}$ one can calculate the quantum mechanical 
version of the spectral function 
\be{2367}
\tilde{C}_{\tbox{qm}}(\omega) =
\sum_{m}|F_{nm}|^2\ 2\pi\delta(\omega - (E_m{-}E_n))
\ee
where it is implicit that the delta functions have  
a finite smearing width related to the measurement time~$t_{\infty}$, 
and an average over the reference state (${E_n \sim E}$) is 
required to reflect the associated uncertainty in energy.

\sect{Correspondence} 
For a chaotic system, if the correlation time is short,  
one expects quantum-to-classical correspondence (QCC) 
with regard to $\tilde{C}(\omega)$. It follows from \Eq{e2367}   
that this spectral function should reflect the bandprofile 
of the perturbation matrix \mycite{mario1,mario2,mario3,mario4}.
Let us express this observation in a convenient way 
that allows a practical procedure for numerical verification. 
We calculate $F_{nm}$ in an energy window 
of interest, and define the associated matrix 
\be{0}
\bm{X} \ \ = \ \ \{|F_{nm}|^2\}
\ee
The bandprofile $\bar{C}_a(r)$ is defined 
by the average of the elements $X_{nm}$ along the 
diagonals $n{-}m=r$.
In the same way we also define a {\em median} based bandprofile $\bar{C}_s(r)$.
Given that the mean level spacing $\Delta_0$ is small compared 
with the energy range of interest,  
the correspondence between $\tilde{C}_{\tbox{qm}}(\omega)$ 
and $\tilde{C}_{\tbox{cl}}(\omega)$ can be expressed as: 
\be{2}
\bar{C}_a(n-m) \ \ \approx \ \ \left(\frac{2\pi}{\Delta_0}\right)^{-1} \ \tilde{C}_{\tbox{cl}}(E_n-E_m) 
\ee
In particular, it follows from \Eq{e13} that the 
unrestricted average value of the elements $X_{nm}$ is 
\be{19}
\langle\langle \bm{X} \rangle\rangle_{\infty} 
\ \ = \ \ \left(\frac{8}{3\pi}\right)  \frac{\mass v_{\tbox{E}}^3}{L_y L_x^2}
\ee
In fact this result can be established without 
relaying on QCC considerations via a sum rule 
that we discuss in \Sec{s6}, and the same 
result is also obtained from the zero order evaluation
of matrix as described in \Sec{s5}. 
Whenever applicable we re-scale the numerical results 
with respect to this reference value.

The applicability of the QCC relation \Eq{e2},  
to the analysis of our billiard system is confirmed 
in \Fig{f33}, down to very small frequencies. We also see that 
\be{0}
\bar{C}_s(r) \ \ \ll \ \ \bar{C}_a(r) \ \ \rightarrow  \ \ \langle\langle \bm{X} \rangle\rangle_{\infty} 
\ee
where the value on the right is obtained in the limit ${r\rightarrow\infty}$.
The inequality $\bar{C}_s \ll \bar{C}_a$  means that the value 
of the typical matrix element is very small compared with the average value. 
We are therefore motivated to define notions of sparsity and texture 
in \Sec{s3}.


\section{Sparsity and texture}

\label{s3}

For strongly chaotic systems the elements within the band 
have approximately a Gaussian distribution. But for WQC 
the matrix becomes {\em sparse} and {\em textured} 
as demonstrated in \Fig{f31}.
These features go beyond the semiclassical analysis of the bandprofile.
The sparsity is related to the size distribution 
of the in-band elements: Loosely speaking one may say that only 
a small fraction ($s$) of elements are large, 
while most of the elements are very small  
(for a precise definition of~$s$ see below).
The texture refers to the non-random arrangement 
of the minority of large elements. 

In the WQC regime the size distribution of 
the in-band elements becomes log-wide (approximately log-normal) 
as seen in \Fig{f32}. 
This is reflected by having ${\bar{C}_s(r) \ll \bar{C}_a(r)}$ 
as seen in \Fig{f33}.
Accordingly, an optional measure for sparsity 
is the parameter~$q$ which is defined 
as the ratio of the median to the average.
  
The sparsity and the texture of $F_{nm}$ are important for the analysis 
of the energy absorption rate as implied by SLRT.
Accordingly, it is physically motivated to characterize the sparsity by 
a resistor network measure~$g_s$ that reflects the connectivity 
of the elements, and hence has a direct relation 
to the semi-linear response characteristics of the system.  
The precise definitions of~$g_s$ is given below.  
For a strictly uniform matrix~${g_s=s=1}$, 
for a Gaussian matrix ${s=1/3}$ and ${g_s\sim1}$, 
while for sparse matrix ${s,g_s\ll1}$. 
The dependence of the sparsity on the energy and on 
the degree of deformation is demonstrated in \Fig{f34}, 
and is related to the mixing of the levels in \Fig{f14}.

\sect{Definition of $s$}
Define a matrix $\bm{X}$ whose elements are~${X_{nm}=|F_{nm}|^2}$. 
Associate with it an untextured matrix $\bm{X}^{\tbox{utx}}$ 
and a uniformized  matrix $\bm{X}^{\tbox{unf}}$ 
that have the same bandprofile~$C(r)$. 
The former is obtained by performing random permutations 
of the elements along the diagonals, 
while the latter is obtained by replacing 
each of the elements of a given diagonal by their average.
The participation number (PN) of a set $\{X_i\}$ is defined as
${\left(\sum_i X_i\right)^2/\sum_i X_i^2}$, 
and reflects the number of the large elements. Here the index is ${i=(n,m)}$.
The PN of $\{X_{nm}\}$ counts the number of large elements in the matrix.
The PN of $\{X_{nm}^{\tbox{unf}}\}$ counts the number of the in-band elements. 
Accordingly, the ratio constitutes a measure for sparsity:
\be{9} 
s \ \ = \ \ s[\bm{X}] \ \ \equiv \ \ \frac{\mbox{PN}\left[ \bm{X} \right]}{\mbox{PN}\left[\bm{X}^{\tbox{unf}}\right]}
\ee
It should be clear that $\bm{X}$ and $\bm{X}^{\tbox{utx}}$
have the same $s$ but only the former might have texture.
So the next question is how to define texture 
avoiding a subjective visual inspection.

\sect{Definition of $g_s$}
Complementary information about the sparsity 
of the matrix, that takes into account the texture as well, 
is provided by the resistor network measure~$g_s$. 
Coming back to $X_{nm}$ we can associate with it a matrix 
\be{28}
G_{nm} = 2F(n{-}m) \, \frac{X_{nm}}{(n-m)^2}
\ee
where $\sum_r F(r)=1$ is a weight function 
whose width should be quantum mechanically large 
(i.e.~$\gg1$) but semiclassically small (i.e.~$\lesssim$~the~bandwidth). 
With such choice the $G_{nm}$ are proportional to the 
Fermi-golden rule transition rates that would be induced 
by a low-frequency driving $-f(t)F$. 
Optionally we can regard these $G_{nm}$ as representing 
connectors in a resistor network, as in \Fig{f10}. 
The inverse resistivity of the strip can be calculated using 
standard procedure, as in electrical engineering, 
and the result we call $\langle\langle \bm{X} \rangle\rangle_s$. 
{For more details} see \App{a0}.
It is useful to notice that if all the elements 
of $\bm{X}$ are identical then $\langle\langle \bm{X} \rangle\rangle_s$ equals the same number.
More generally $\langle\langle \bm{X} \rangle\rangle_s$ 
is smaller than the conventional algebraic average $\langle\langle \bm{X} \rangle\rangle_a$ 
(calculated with the same weight function). 
Accordingly, the resistor network quantity $\langle\langle \bm{X} \rangle\rangle_s$  
can be regarded as a smart average over the elements of~$\bm{X}$,  
that takes their connectivity into account. 
Consequently, it is natural to define a physically motivated 
resistor-network measure for sparsity and texture:   
\be{29} 
g_s \ \ = \ \ g_s[\bm{X}] \ \ \equiv \ \ \frac{\langle\langle \bm{X} \rangle\rangle_s}{\langle\langle \bm{X} \rangle\rangle_a}
\ee
One can show that $\langle\langle \bm{X} \rangle\rangle_s$ 
is strictly bounded from below by the {\em harmonic} average. 
In practice the {\em geometric} average or the {\em median} provide better lower bounds.
In the RMT context a realistic estimate for $\langle\langle \bm{X} \rangle\rangle_s$ can 
be obtained using a generalized variable-range-hopping procedure (see~\mycite{kbd} for details).

\sect{Additional definitions}
If the elements $X_{nm}$ have a well defined 
average $\langle\langle \bm{X} \rangle\rangle_{\infty}$
in the limit of infinite truncation, then it is 
convenient to define   
\be{300} 
g_c \ \  &\equiv& \ \ \frac{\langle\langle \bm{X} \rangle\rangle_a}{\langle\langle \bm{X} \rangle\rangle_{\infty}}
\\ \label{e310}
g \ \  &\equiv& \ \ \frac{\langle\langle \bm{X} \rangle\rangle_s}{\langle\langle \bm{X} \rangle\rangle_{\infty}}
\ \ = \ \ g_s \ g_c
\ee
Later, in \Sec{s8}, we discuss the physical significance of~$g$, 
and identify it as the dimensionless 
absorption coefficient. In particular, $g_c$ is identified as 
the dimensionless absorption coefficient in the classical calculation, 
which is determined by taking into account classical correlations.  
In the quantum case we have an additional suppression factor~$g_s$ 
due to the sparsity of the perturbation matrix.

\section{Classical analysis of $\tilde{C}(\omega)$}

\label{s4}

Let us assume that we have a collision in angle~$\theta$ 
with the flat piston. The force which is exerted on the particle 
during the collision is $F_0$, such that the impact is 
\be{0}
q_{\theta} =  2\mass v_{\tbox{E}} \cos(\theta)
\ee
Consequently, the force $F(t)$ looks like a train of spikes as in \Eq{e6}. 
Note that the duration of a collision is $q_{\theta}/F_0$. 
In the absence of deformation the time distance between the spikes is 
\be{0}
\tau_{\theta} = \frac{2L_x}{v_{\tbox{E}}\cos\theta}
\ee
where $\theta$ is constant of the motion.  
For the following calculations it is useful to define 
the following averages: 
\be{377}
C_{\infty} 
= \left\langle\left(\frac{q_{\theta}^2}{\tau_{\theta}}\right)\right\rangle_{\theta}
&=& \frac{8}{3\pi} \frac{\mass^2 v_{\tbox{E}}^3}{L_x} 
\\
\mylabel{e30}
c_0 =
\left\langle \left(\frac{q_{\theta}}{\tau_{\theta}}\right)^2 \right\rangle_{\theta}
&=& \frac{3}{8} \frac{\mass^2 v_{\tbox{E}}^4}{L_x^2}
\\
\mylabel{e31}
c_{\infty} = 
\left\langle\left(
\frac{q_{\theta}}{\tau_{\theta}}\right)\right\rangle_{\theta}^2
&=& \frac{1}{4} \frac{\mass^2 v_{\tbox{E}}^4}{L_x^2}
\\
\mbox{Var}\!\left[\frac{q}{\tau}\right] 
\equiv
c_0-c_{\infty} 
&=& \frac{1}{8} \frac{\mass^2 v_{\tbox{E}}^4}{L_x^2}
\ee

If we have a very small~$u$ the effect 
would be to ergodize~$\theta$ with some rate~$\gamma_{\theta}$.
After time $t$ the number ($\#$) of collisions is ${t/\tau_{\theta}}$ 
and consequently the deviation of the perturbed trajectory 
is multiplied by  $(1+L/R)^\# \approx \exp\left((v_{\tbox{E}}t/R)\cos(\theta)\right)$.  
Accordingly, the instability exponent is 
\be{33}
\gamma_{\theta} \ \approx \ \gamma_0 + \frac{v_{\tbox{E}}}{R} \cos(\theta)
\ee 
For sake of generality we have added a background term $\gamma_0$.
This background term would arise if the upper or lower walls were deformed, 
or if the potential floor were not flat. 
A non zero $\gamma_0$ is unavoidable in a realistic system. 
As we shall see shortly the effect of the deformation 
is twofold. The primary effect is to ergodize~$\theta$,  
and the secondary effect is to modify the small~$\omega$ 
spectral content of the fluctuations.

\sect{Power spectrum}
Let us define $F_{\theta}(t)$ as the temporal ``signal" 
which is associated with a trajectory 
that {\em starts} at the piston with $\theta$ collision angle. 
This signal consists of delta spikes,  
the first one being $q_{\theta}\delta(t)$.  
The correlation function can be expressed as  
\be{34}
\langle F(0)F(t) \rangle
= \left\langle \frac{q_{\theta}}{\tau _{\theta}} F_{\theta}(t) \right\rangle_{\theta}
\ \ \ \ \ \ \ \ \ \ \\ \label{e355}
= \left\langle \frac{q_{\theta}^2}{\tau _{\theta}} \right\rangle_{\theta} \delta(t) 
\ + \ \mbox{correlations}
\ee
where the first term represents the self-correlation of the spikes. 
It is convenient to subtract from $\langle F(0)F(t) \rangle$ 
its global offset, and to define the correlation function as 
\be{0}
C(t) \ \ = \ \ \langle F(0)F(t) \rangle - \langle F \rangle^2
\ee
The associated power spectrum is the Fourier transform:
\be{35}
\tilde{C}(\omega)
= \left\langle \frac{q_{\theta}}{\tau _{\theta}} \tilde{F}_{\theta}(\omega) \right\rangle_{\theta}
- \left\langle \frac{q_{\theta}}{\tau _{\theta}}  \right\rangle_{\theta}^2 2\pi\delta(\omega)
\ee
Note that $\tilde{F}_{\theta}(\omega)$ is the FT of $F_{\theta}(t)$.
It is not the same as $\tilde{F}(\omega)$ of \Eq{e11}.
The latter has a random phase due to a time displacement of the time origin, 
while in the case of $F_{\theta}(t)$ the time origin 
is fixed by the presence of $F(0)$ in \Eq{e34}. 

\sect{Infinite frequency limit}
The first obvious observation is that that for large frequencies 
the power spectrum becomes flat and reaches a constant value
that reflects the self-correlation peak of \Eq{e355} 
\be{271}
\tilde{C}(\infty) 
\ \ = \ \ \left\langle\frac{q^2}{\tau}\right\rangle
\ \ = \ \ C_{\infty}
\ee
where $C_{\infty}$ is given by \Eq{e377}.
This result, if it is a applied to finite frequencies, 
is termed in the literature ``the white noise approximation".

\sect{Zero deformation} 
Let us consider the non-deformed integrable billiard.  
Then the bouncing trajectories consist of equal spikes 
and may have an arbitrary long periods $\tau_{\theta}$.
The Fourier transform of ${F_{\theta}(t)=\sum_j q_{\theta}\delta(t{-}t_j)}$, 
is a reciprocal comb, namely 
\be{0}
\tilde{F}_{\theta}(\omega) = q_{\theta}
\sum_{n} \frac{2\pi}{\tau_{\theta}} \delta\left(\omega -
\frac{2\pi}{\tau_{\theta}}n \right)
\ee
The power spectrum is obtained using  \Eq{e35}.
It consists of two components. 
One component is the zero frequency peak 
which reflects the dispersion of the impact pulses  
\be{43}
\tilde{C}(\omega\sim0) 
\ \ = \ \
\mbox{Var}\!\left[\frac{q}{\tau}\right]\ 2\pi\delta(\omega) 
\ee
This zero frequency peak would be broadened 
if the deformation were non zero,
as discussed in the next paragraph. 
The second component of the power spectrum 
consists of ballistic peaks at ${\omega_n=(\pi v_{\tbox{E}}/L_x)n}$, 
that merge to $C_{\infty}$ in the infinite frequency limit:
\be{10}
\tilde{C}(\omega > 0) \ \ = \ \ 
C_{\infty} 
\sum_{\omega_n > \omega} 
\frac{3}{2n} 
\frac{(\omega/\omega_n)^4}{\sqrt{1 - (\omega/\omega_n)^2}}
\ee
\Fig{f33} presents the numerical data for a slightly 
deformed billiard. Disregarding the broadened 
zero-frequency peak, the above zero-deformation result 
provides a practical overall approximation.

\sect{Small deformation} 
For small deformation the main effect is the broadening 
of the delta function in \Eq{e43}. Assuming a $\theta$ 
independent $\gamma$, the $\delta(\omega)$ is replaced  
by the Lorentzian $(1/\pi)\gamma/(\omega^2+\gamma^2)$.
Hence, we get for small frequencies
\be{46}
\tilde C(\omega\ll\Delta_{\tbox{L}}) \ \ \approx \ \  
\mbox{Var}\!\left[\frac{q}{\tau}\right] \times 
\frac{2/\gamma}{1+(\omega/\gamma)^2}
\ee
We further illuminate this result using a time 
dependent and number variance approaches in \App{a1}. 
If ${\gamma=\gamma_0}$ is well defined there is 
a well defined limiting value as ${\omega\rightarrow 0}$. 
With the identification $\gamma\sim 1/t_{\tbox{R}}$ 
one should realize that the power spectrum at zero frequency 
is enhanced by factor $t_{\tbox{R}}/t_{\tbox{L}}$, hence
\be{42}
\tilde{C}(\omega = 0) \ \ \approx \ \ \frac{1}{u} C_{\infty}
\ee
But if ${\gamma}$ is given by \Eq{e33} we have to 
perform an ergodic average over $\theta$. 
This becomes interesting if $\gamma_0$ is very small 
or zero, as discussed in the next paragraph.

\sect{Bouncing effect}
It has been proven \mycite{bunimovich,young} that in strictly hyperbolic billiards 
the time correlation function exhibits an exponential decay rate. 
But if there are bouncing trajectories, that do not 
collide with the deformed surfaces, and can be of arbitrarily long length, 
then a power law decay shows up as in the hard-sphere gas \mycite{bouncing1} 
and in the Stadium \mycite{bouncing2}. Our billiard, 
with  ${\gamma}$ as given by \Eq{e33},  
can be regarded as a related variation on this theme. 
Assuming that $\gamma_0$ is very small,  
the trajectory has a very long bouncing period
when $\theta \sim \pi/2$.  
Consequently, the ergodic average over $1/\gamma$ 
generates a logarithm factor $\log(1/\gamma_0)$, 
or at finite frequency it becomes $\log(1/\omega)$.
Let us be more precise in the $\gamma_0=0$ case.
Averaging over the Lorentzian we get  
\be{-1}
\tilde C(\omega) = 
\frac{\mass^2  v_{\tbox{E}}^3}{2 L_x^2}
\frac{R}{\sqrt{1 + \frac{\omega^2 R^2}{v_{\tbox{E}}^2}}}
\ \mbox{arctanh} 
\left[\frac{1}{\sqrt{1 + \frac{\omega^2 R^2}{v_{\tbox{E}}^2}}}\right] 
\\ \ \ \label{e480}
= \frac{\mass^2  v_{\tbox{E}}^4}{4L_x^2}
\frac{t_{\tbox{R}}}{\sqrt{1 + \omega^2 t_{\tbox{R}}^2}}
\ln\left[ 1 + \frac{2+2\sqrt{1 + \omega^2 t_{\tbox{R}}^2} }
{\omega^2 t_{\tbox{R}}^2} 
\right]
\ee
For small $\omega$ the above expression can be further simplified:
\be{27}
\tilde C(\omega\ll\Delta_{\tbox{R}}) \ \ \approx \ \ 
\mass^2  v_{\tbox{E}}^3 \ 
\frac{R}{2L_x^2} \ 
\ln \left[\frac{2}{\omega t_{\tbox{R}}}\right]
\ee
In the zero frequency limit, if $\gamma_0$ is finite, 
the logarithmic factor in the baove expression 
is replaced by $\ln[2/(\gamma_0 t_{\tbox{R}})]$. 
Consequently, the result $g_c=1/u$ which is implied 
by \Eq{e42} is replaced by 
\be{0}
g_c 
\ = \ \ln\left[2\frac{\Delta_{\tbox{R}}}{\gamma_0}\right] \ \frac{1}{u} 
\ \ \ \ \ \ \mbox{[for $\omega_c\rightarrow 0$]}
\ee
In the quantum case, that we discuss later, the finite 
level spacing provides an additional lower cutoff $\Delta_0$ 
that ``competes" with $\gamma_0$ as discussed in \Sec{s6}.

\section{Perturbation theory analysis}

\label{s5}

In this section we shall see what comes out for the matrix elements~$F_{nm}$
within the framework of quantum perturbation theory to leading order:
zero order evaluation for the ``large" elements, 
and first order perturbation theory (FOPT) for the ``small" elements. 
In \Sec{s6} we shall try to reconcile the perturbation theory 
results with the classical results of \Sec{s4}.

The small parameter in the perturbative treatment is~$u$. 
The eigenstates ${\bar{\bm{n}}=(n_x,n_y)}$ of the non-deformed billiard are
\be{0}
\psi^{(\bar{\bm{n}})}(x,y) = 
\frac{2}{\sqrt{L_x L_y}}\sin\left(n_x\frac{\pi}{L_x} x\right) \sin\left(n_y\frac{\pi}{L_y} y\right) 
\ee
The deformation profile is 
\be{0}
D_u(y) &=& \sqrt{R^2-(y{-}\varepsilon)^2} 
- \sqrt{R^2-(L_y{-}\varepsilon)^2} 
\ee
In the FOPT treatment the perturbation term in the Hamiltonian 
is calculated using an expression analogous to \Eq{e16}, 
with $D_u$ replacing $D_f$ along the left wall, leading to 
\be{0}
U_{\bar{\bm{n}} \bar{\bm{m}}} = -\frac{\pi}{\mass L_x^3} \left(D_{n_y-m_y}-D_{n_y+m_y}\right) n_x m_x
\ee
where 
\be{0}
D_{\nu} \equiv \frac{1}{L_y}\int_0^{L_y} D_u(y) \cos\left(\nu\frac{\pi}{L_y} y\right) dy
\ee
In the numerical analysis we calculate $D_{\nu}$ 
and hence $U_{\bm n \bm m}$ numerically. 
But here, for presentation purpose, 
we introduce a practical approximation: 
\be{0}
|U_{\bm n \bm m}| \approx \left(\frac{D_0}{\mass L_x^3}\right) \frac{n_x m_x}{1+|n_y-m_y|^{\alpha}}
\ee
In this expression an exponent ${\alpha{=}1}$  
would arise due to the discontinuity of $D_u(y)$ at ${y=0}$.
However, the effective value of $\alpha$ is larger because 
this discontinuity is very small and hardly expressed numerically.
Furthermore, we would not like to restrict the analysis 
to the specific deformation that had been assumed in the numerics.
We therefore regard $\alpha$, for the sake of further 
discussion, as a fitting parameter.  

We can regard the deformation $U$ as inducing scattering 
between the $n_y$ modes of the rectangular ``waveguide".
If the box is not deformed ($u{=}0$), which is like ``no scattering'',   
then $n_y$ is a good quantum number. 
Otherwise, for non-zero deformation, the levels are mixed.  
The FOPT overlap between perturbed and unperturbed states is
\be{39}
\langle \bar{\bm{m}} | n \rangle  
= \frac{U_{\bar{\bm{m}}\bar{\bm{n}}}}{E_{\bar{\bm{n}}} - E_{\bar{\bm{m}}}}
\ee
Note that by adiabatic continuation we assume in this expression 
an association of perturbed states~$n$ with unperturbed 
states~${\bar{\bm{n}}=(n_y,n_y)}$. This association holds for  
those levels that are not mixed non-perturbatively. 
Later we discuss the coexistence of perturbative and non-perturbative mixing.

\sect{Zero order elements}
We turn to look at $F_{nm}$. For zero deformation 
it is block-diagonal with respect to~$n_y$. Namely, 
\be{47}
F_{\bar{\bm{n}}\bar{\bm{m}}} 
\ \ = \ \ 
-\delta_{n_y, m_y} \ \frac{\pi^2}{\mass L_x^3} n_x m_x
\ee
Most of the matrix elements are zero, while  
a small fraction are finite. Considering the elements 
within an energy shell~$E$, 
setting ${|n| \sim |m| \sim k_{\tbox{E}} L}$, 
the size of the large elements is   
\be{48}
|F_{nm}|_{0}
\sim \frac{1}{\mass L^3}  (k_{\tbox{E}} L)^2 
\ee
In \App{a2} we show that the fraction 
of elements that have this large value is 
\be{0}
p_0 =\frac{2}{\pi k_{\tbox{E}} L_y}
\ee
Consequently, the average value $\langle\langle |F_{nm}|^2 \rangle\rangle_{\infty}$    
of the elements is $p_0 \times |F_{nm}|_{0}^2$. 
In the more careful calculation of \App{a2} we show that   
\be{0}
\langle\langle |F_{nm}|^2 \rangle\rangle_{\infty} 
\ = \ 
\frac{8}{3\pi}\frac{k_{\tbox{E}}^3}{\mass^2 L_x^2L_y}
\ee
in consistency with the semiclassical relation \Eq{e2}.

\sect{FOPT elements}
For small~$u$ the large size matrix elements
of $F_{nm}$ are hardly affected by the mixing. 
But at the same time the deformation gives rise 
to in-band small size matrix elements, 
that would have been zero if~$u$ were zero. 
Within FOPT the following approximation applies:
\be{0}
F_{nm} &=& 
\sum_{n',m'} 
\langle n| \bar{\bm{n}}' \rangle 
F_{\bar{\bm{n}}'\bar{\bm{m}}'}
\langle \bar{\bm{m}}' | m\rangle      
\\
&\approx& 
F_{\bar{\bm{n}}\bar{\bm{m}}}
+ \bra \bar{\bm{n}} |m\ket F_{\bar{\bm{n}}\bar{\bm{n}}}
+ \bra \bar{\bm{m}} |n\ket^* F_{\bar{\bm{m}}\bar{\bm{m}}} \\
&=&
F_{\bar{\bm{n}}\bar{\bm{m}}} + \bra \bar{\bm{n}} |m\ket 
(F_{\bar{\bm{n}}\bar{\bm{n}}} - F_{\bar{\bm{m}}\bar{\bm{m}}})
\ee
Hence, the emerging small elements are
\be{0}
&& |F_{nm}|  =
\left|\frac{U_{\bar{\bm{n}}\bar{\bm{m}}}}{E_{\bar{\bm{n}}} - E_{\bar{\bm{m}}}}\right|
|F_{\bar{\bm{n}}\bar{\bm{n}}} - F_{\bar{\bm{m}}\bar{\bm{m}}}| 
\\ 
&& \ \ \approx  \frac{D_0}{\mass L^4} 
\frac{\left(n_x^2-m_x^2\right)n_x m_x}{\left(|n|^2-|m|^2\right)\left(1+|n_y-m_y|^{\alpha}\right)}
\ \ \ \ 
\ee
where for simplicity we had assumed $L_x\sim L_y\sim L$ 
such that  ${E_n\approx\pi^2 |n|^2/(2\mass L^2)}$ 
with ${|n|\equiv (n_x^2+n_y^2)^{1/2}}$.

Given an energy window around~$E$, 
we would like to estimate the typical size 
of the elements $F_{nm}$ that connect 
energy levels that have the 
separation ${|E_n-E_m| = \omega}$.
Our interest is in small frequencies ${\Delta_0 \ll \omega \ll \Delta_{\tbox{L}}}$.
Setting ${||n|-|m|| \sim \omega/\Delta_{\tbox{L}}}$,  
and ${D_0\approx L^2/R}$, 
and ${|n_x{-}m_x| \sim |n_y{-}m_y| \sim |n| \sim |m| \sim k_{\tbox{E}} L}$, 
we get for the majority of elements the estimate 
\be{71}
|F_{nm}|_{\tbox{FOPT}} \ \sim \  
\left(\frac{\Delta_{\tbox{R}}}{\omega}\right) \times
\frac{1}{\mass L^3} (k_{\tbox{E}} L)^{3-\alpha}  
\ee
This should be contrasted with the zero order 
value \Eq{e48} of the large but rare elements: 
it is much smaller whenever the FOPT estimate applies.

\section{Quantum analysis of $\tilde{C}(\omega)$}

\label{s6}

The QCC relation \Eq{e2} implies that $\tilde{C}_{\tbox{qm}}(\omega)$  
reflects the algebraic average over the elements of the matrix $\{ |F_{nm}|^2 \}$
along the diagonal $|E_n-E_m|\sim \omega$. 
Our numerics show that we can trust \Eq{e2} up to the very small frequency~$\Delta_0$. 
This statement is based on some assumptions that should be clarified.

First we would like to emphasize that both classically and quantum mechanically
\be{77}
\tilde{C}_{\tbox{qm}}(\omega \gg \Delta_{\tbox{L}}) \ \ \approx \ \  C_{\infty}
\ee
In the classical context this value merely reflects the self correlation  
of the spikes of which $F(t)$ consists, and hence it is proportional 
to the ratio between the area (length) of the piston and the volume (area) of the box (billiard). 
In the quantum context it reflects the associated assumption that 
well separated eigenstates look like uncorrelated random waves, 
and hence $|F_{nm}|^2$ is determined by the same ratio as in the classical case.
For more details see appendices of \mycite{frc}.   

\sect{QCC condition}
As we go to smaller frequencies, correlations on larger time scales 
become important, and the validity of the QCC relation \Eq{e2} 
becomes less obvious. Recall that due to the bouncing 
\be{771}
\tilde{C}_{\tbox{cl}}(\omega {\sim} 0) \ \ \approx \ \ \frac{1}{u}  C_{\infty}
\ee
Recall also that the matrix elements are strictly bounded from above. 
The maximal value is in fact given by \Eq{e48} and accordingly  
\be{0}
\tilde{C}_{\tbox{qm}}(\omega) \ \ < \ \ \frac{1}{\hbarr} C_{\infty}
\ee
This has an immediate implication:  
QCC cannot hold globally unless $\hbarr<u$. 
This requirement can be illuminated 
from an optional perspective.
The zero frequency peak of $\tilde{C}_{\tbox{cl}}(\omega)$ 
has a width $\Delta_{\tbox{R}}$. 
This peak cannot be resolved by $\tilde{C}_{\tbox{qm}}(\omega)$ 
unless ${\Delta_{\tbox{R}}>\Delta_0}$. 
Again we get the same necessary condition
\be{80}
\hbarr \ \ < \ \ u,  
\ \ \ \ \ \ \ \ \ \mbox{[QCC requirement]}
\ee

\sect{Sum rule}
Extending the discussion with regard to \Eq{e77}, 
it is important to realize that the integral over $\tilde{C}_{\tbox{cl}}(\omega)$ 
equals $\mbox{Var}(F)$, and accordingly it does not depend on~$u$, 
but only on the ratio between the area (length) of the piston
and the volume (area) of the box (billiard).  
Note that the height of the zero frequency peak 
is proportional to~$1/u$, while its width is proportional to~$u$ 
in consistency with this observation.

In complete analogy, in the quantum analysis the sum $\sum_m |F_{nm}|^2$ 
does not depend on~$u$. If the diagonal elements can be neglected it follows 
that  $\tilde{C}_{\tbox{qm}}(\omega)$ does not depend on~$u$. But if $\hbarr>u$,  
the zero frequency peak cannot be resolved, and the deficiency can be 
attributed to the diagonal elements, in consistency with the FOPT analysis.

\sect{FOPT}
In the regime $u<\hbarr$ it is instructive to contrast 
the lower bound FOPT result which is implied by \Eq{e71}, 
with the SC result which is implied by \Eq{e46}
\be{81}
\tilde{C}_{\tbox{qm-FOPT}}(\omega) 
& \sim & \left(\frac{1}{\hbarr}\right)^{3-2\alpha}
\frac{\Delta_{\tbox{R}}^2}{\omega^2+\Delta_0^2} \ C_{\infty}
\\ \label{e82}
\tilde{C}_{\tbox{qm-SC}}(\omega) 
&\approx& 
\frac{\Delta_{\tbox{L}} \Delta_{\tbox{R}}}{\omega^2+\Delta_{\tbox{R}}^2} \ C_{\infty}
\ee
The lower cutoff $\Delta_0$ in the FOPT expression 
has been entered by hand to indicate its existence.  
It is implicit here that the frequency range of interest 
is $\Delta_{\tbox{R}}\ll\omega\ll \Delta_{\tbox{L}}$. 
In the worst case of having a deformation with 
discontinuity (${\alpha=1}$), the ratio between these 
two results, in the frequency range of interest, 
is as one could expect ${(u/\hbarr) \ll 1}$.
We shall discuss the relevance of the FOPT and semiclassical 
expressions below, and also in \Sec{s7}.

\sect{Evaluation of $g_c$}
Coming back to the regime ${\hbarr<u}$, 
assuming that the QCC relation \Eq{e2} can be trusted,  
we deduce that the unrestricted average value 
of the matrix elements at energy~$E$ is 
\be{0}
\langle\langle |F_{nm}|^2 \rangle\rangle_{\infty} 
\ = \ 
\left(\frac{2\pi}{\Delta_0}\right)^{-1} C_{\infty}
\ee
Our interest is in the response characteristics of the system for low frequency driving, 
which we further discuss later in \Sec{s8}.
We assume that the spectral content of the driving is characterized by a 
cutoff frequency ${\Delta_{\tbox{R}}  < \omega_c < \Delta_{\tbox{L}}}$. 
Therefore we look on the band-averaged value:
\be{0}
\langle\langle |F_{nm}|^2 \rangle\rangle_a 
\ \equiv \   
\left(\frac{2\pi}{\Delta_0}\right)^{-1} 
\,\frac{1}{\omega_c} 
\int_{\Delta_0}^{\omega_c} \tilde{C}_{\tbox{qm}}(\omega) d\omega
\\
\ \ \ \ \ \ \ \ \equiv \ 
g_c 
\times \langle\langle |F_{nm}|^2 \rangle\rangle_{\infty}    
\ee
If QCC holds, and $\Delta_0$ is taken to be zero, 
then we should get the classical result: 
in accordance with the ``sum rule" the expected enhancement 
factor would be ${g_c\approx 1}$ if ${\omega_c\sim\Delta_{\tbox{L}}}$, 
and ${g_c \approx \Delta_{\tbox{L}}/\Delta_{\tbox{R}}}$ 
if ${\omega_c\sim\Delta_{\tbox{R}}}$.
But $\Delta_0$ is finite, and we get
\be{86}
g_c \mbox{[qm]}
\ \approx \  
\left[1-
\frac{\Delta_0}{\Delta_{\tbox{R}}} 
\ln\left(2\frac{\Delta_{\tbox{R}}}{\Delta_0}\right)
\right]
\times g_c \mbox{[cl]}
\ee
which is analogous to ``weak localization corrections" 
to the mesoscopic conductance of closed rings \mycite{kamenev}.

\sect{Evaluation of $g_s$}
The typical value of the elements, unlike the average value, 
is dominated by the majority of small elements. 
In order to calculates $g_s$ as defined in \Eq{e29},
we have to bridge between the FOPT and the semiclassical
analysis. To do it in a mathematically rigorous way seems 
to be impossible. We therefore extend standard phenomenology
and test it against numerical results. 
The basic idea is that FOPT cannot be trusted globally 
once levels are mixed non-perturbatively, 
but still it can be used in a restricted way. 
The analogy here is with Wigner's Lorentzian 
whose {\em tails} are given correctly by FOPT, 
in-spite of the non-perturbative mixing of levels.
See discussion of this issue in~\mycite{frc}.  

It is natural to expect FOPT to hold as an estimate 
for the majority of {\em small} elements as 
long as it does not exceed the semiclassical estimate.
If we take a band matching cutoff ${\omega_c\sim\Delta_{\tbox{R}}}$, 
and calculate the ratio of the ``area" under \Eq{e81} 
to the ``area" under \Eq{e82} we get:
\be{0}
g_s 
\ \approx \ 
\left(\frac{1}{\hbarr}\right)^{6-4\alpha} u^2
\ee
Note that with $\alpha=1$ it follows 
that $g_s \propto (u/\hbarr)^2$. 
In our numerics we fix $\omega_c$ 
as the first minimum of $C_a(\omega)$ 
implying ${\omega_c\sim\Delta_{\tbox{L}}}$, 
and consequently ${g_c\sim1}$, and ${g\sim g_s}$.
Our numerics fits well to ${g \propto u^2/\hbarr}$, 
indicating that the effective $\alpha$ 
is somewhat larger than unity.

At this point one should appreciate how the 
contradicting FOPT and semiclassical results 
reconcile. The former apply to the majority 
of elements while the latter apply to the 
algebraic average which is dominated by relatively  
rare elements. The WQC regime where this picture is 
valid is further discussed in \Sec{s7}.

For completeness one should be aware that 
the {\em typical} (median) value of the elements in $\bm{X}$ 
provides an underestimate for the 
resister network average $\langle \langle\bm{X}\rangle \rangle_s$.
The reason is very simple: even if the matrix is very sparse (${s\ll1}$) 
a network becomes {\em percolating} if the bandwidth 
is large enough. An RMT perspective~\mycite{kbd}, 
that uses a generalized variable-range-hopping approach, 
implies the following prescription: 
\be{0}
g \ \ \mapsto \ \ \max\left\{1, \ g \exp\left[\sqrt{-\ln b \ln g} \right]\right\}
\ee
Here ${b=\omega_c/\Delta_0}$ is the dimensionless bandwidth. 
This prescription allows to ``correct" the result 
that has been deduced for~$g$ on the basis of a typical value estimate 
of the matrix elements. It is required if~$b$ is large.

\section{The WQC regime}

\label{s7}

Quantum mechanics introduces in the billiard problem an 
additional frequency scale $\Delta_0$ that corresponds 
to the mean level spacing.  We can associate with it 
the Heisenberg time $t_{\tbox{H}}=2\pi/\Delta_0$.
It is also possible to define the Ehernfest time $t_{\tbox{E}}$ 
which is required for the exponential instability 
to show up in the quantum dynamics. One can write 
\be{0}
t_{\tbox{H}} &=& (1/\hbarr)^{d{-}1}t_{\tbox{L}} \\
t_{\tbox{E}} &=& [\log(1/\hbarr)] t_{\tbox{R}}
\ee
where $d{=}2$. 
The traditional condition for ``quantum chaos"  is ${t_{\tbox{E}} \ll t_{\tbox{H}}}$, 
but if we neglect the log factor it is simply ${t_{\tbox{R}} \ll t_{\tbox{H}}}$. 
This can be rewritten as ${\Delta_{\tbox{R}} \gg \Delta_0}$, 
which we call the frequency domain version of the quantum chaos condition. 
Optionally one may write a {\em parametric version} 
of the quantum chaos condition, namely ${u \gg u_b}$, where 
\be{0}
u_b = \hbarr  \ \ \ \ \ \ \ \mbox{[de~Broglie deformation]}
\ee
Note that it is the same as the QCC requirement of \Eq{e80}.
Namely, the frequency domain version of this condition 
implies that it should be possible to resolve the zero frequency 
peak of $\tilde{C}(\omega)$ as in \Fig{f33}, 
while the parametric version means 
that a de~Broglie wavelength deformation of the boundary  
is required to achieve ``Quantum chaos".

In practice we witness a WQC regime instead of hard chaos.
We observe in the upper panel of \Fig{f34}  
that ${g_s}$ is significantly smaller than unity, 
even for very small values of~$\hbarr$ 
for which ${u > u_b}$ is definitely satisfied.  
For completeness we show in the lower plot additional 
data points in the regime ${u < u_b}$ where this breakdown 
of QCC is not a big surprise. 
We conclude that QCC for ${u > u_b}$ is restricted to $\tilde{C}(\omega)$, 
and does not imply {\em Hard} quantum chaos (HQC), 
but only WQC. In the WQC regime ${\bar{C}_s(r)\ll \bar{C}_a(r)}$
and consequently ${g_s \ll 1}$, indicating sparsity.

This emergence of the WQC regime can be explained 
by extrapolating FOPT considerations.  
If a wall of a billiard is deformed, the levels are mixed.
FOPT is valid provided  ${|U_{nm}| < \Delta_0}$. 
This condition determines a parametric scale~$u_c$. 
If the unperturbed billiard were chaotic, the variation 
required for level mixing would be~\mycite{prm} 
\be{0}
u_c \approx \hbarr/(k_{\tbox{E}}L)^{1/2} = \hbarr^{3/2}
\ \ \ \ \ \ \mbox{[not applicable]}
\ee
This expression assumes that the eigenstates look like random waves.  
In the Wigner regime (${u_c<u<u_b}$) there is 
a Lorentzian mixing of the levels and accordingly
\be{0}
\mbox{\# mixed levels} \ \approx \ (u/u_c)^2
\ \ \ \ \ \ \mbox{[not applicable]}
\ee
But our unperturbed (rectangular) billiard is not chaotic, 
the unperturbed levels of the non-deformed billiards
are not like random waves.  
Therefore, the mixing of the levels is {\em non-uniform}.
\Fig{f14} illustrates the mixing vs~$u$.

By inspection of the $U_{n_xn_y,m_xm_y}$ matrix elements 
one observes that the dominant matrix elements that are responsible 
for the mixing are those with large~${n_x}$ but small ${|n_y{-}m_y|}$. 
Accordingly, within the energy shell ${E_{n_xn_y} \sim E}$, 
the levels that are mixed first are those with maximal ${n_x}$, 
while those with minimal ${n_x}$ are mixed last. 
The mixing threshold for the former is 
\be{0}
u_c \ \ \approx \ \ \hbarr/(k_{\tbox{E}}L) \ \ = \ \ \hbarr^{2}
\ee
while for the latter one finds  $u_c^{\infty} \sim \hbarr^{0}$, 
which is much larger than $u_b=\hbarr^{1}$. 
In our numerics $g\approx u^2/\hbarr$, implying that the WQC-HQC crossover is at 
\be{325}
u_s \ \ = \ \ \hbarr^{1/2}
\ee
and not at ${u_b=\hbarr}$.    
Accordingly, the WQC regime extends well beyond the 
traditional boundary of the Wigner regime, 
and in any case it is well beyond the FOPT border~$u_c$.

\sect{WQC in broader perspective } 
In a broader perspective the term WQC is possibly 
appropriate also to system with zero Lyapunov exponent ($t_{\tbox{R}}{=}\infty$), 
e.g. the triangular billiard~\mycite{triang}, 
and pseudointegrable billiards~\mycite{spectral}, 
and to systems with a classical mixed phase space. 
But in the present study we wanted to consider 
a globally chaotic system, under semiclassical circumstances such 
that $\Delta_{\tbox{R}}$ is quantum mechanically resolved and QCC is naively expected. 
In this context there are of course other interesting aspects, 
such as bouncing related corrections to Weyl's law~\mycite{Backer}, 
and non-universal spectral statistics issues (see below), 
while our interest was with regard to the semi-linear response 
characteristics of the system.

\sect{Spectral statistics in the WQC regime  } 
The spectral statistics in the WQC regime has been 
studied in~\mycite{wqc1} concerning nearly circular stadium billiard, 
and in~\mycite{wqc2} concerning circular billiards with a rough boundary. 
The model that we analyze is not identical, but can be regarded 
as a variation on the same theme. In \Fig{f21} we display 
some results for the level spacing statistics $P(S)$, 
where the statistics is over ${S_n = (E_{n+1}-E_n)/\Delta_0}$. 
It can be fitted to the cumulative Brody distribution
\be{0}
F(S) = 1 - \eexp{-b S^{q+1}}, 
\ \ \ \ \ 
b = \left[\Gamma\left(\frac{q+2}{q+1}\right)\right]^{q+1}
\ee
which interpolates between the Poisson distribution ($q{=}0$) 
and with the Wigner surmise ($q{=}1$).
This cumulative distributions can be transformed 
into linear functions ${T(x) = \ln\left[-\ln(1-F(\eexp{x}))\right]}$ 
with respect to the variable $x=\ln(S)$, 
and the fitting to our data gives $q{=}0.38$.

Let us remind very briefly how the WQC border is determined 
in this context. It is convenient to describe the dynamics 
using a Poincare map, which relates the angle $\theta_{\tau}$  
of successive collisions (${\tau=1,2,3,\cdots}$) with the piston.
One observes that due to the accumulated effect of 
collisions with the deformed boundary, there is a slow diffusion  
of the angle with coefficient 
\be{0}
D_{\theta} \ \sim \ u^2
\ee
Accordingly, the classical ergodic time is 
\be{0}
\tau_{u} \ \sim \ 1/D_{\theta} \ \sim \ 1/u^2
\ee
and the quantum breaktime due to a dynamical localization 
effect is 
\be{0}
\tau_{h} \ \sim \ D_{\theta}/\hbarr^2 \ \sim \ (u/h)^2
\ee
The border of the WQC regime is defined by the 
condition ${\tau_{h}<\tau_{u}}$ leading to~\Eq{e325}.
However, we would not like to over-emphasize this consistency 
because it is not a-priori clear that spectral-statistics
and sparsity related characteristics always coincide.

\sect{Intensity statistics in the WQC regime  } 
WQC is also reflected in the intensity statistics 
of the wavefunctions. If we had HQC we would expect 
Porter-Thomas (Gaussian) statistics and random wave  
correlations. The wavefunctions that we find do not 
look like random waves.  In \Fig{f22} we show the 
statistics of the integrated intensity:
\be{100}
I_n \ = \ \frac{1}{2k_n^2}
\int_0^{L_y} |\varphi^{(n)}(y)|^2 dy 
\ = \ -\frac{1}{2 E_n}F_{nn}
\ee
Note that the total intensity, which is obtained 
by integrating along the whole boundary with proper weight, 
gives unity, corresponding to the normalization of the wavefunction.

\section{The heating rate problem} 

\label{s8}

In this section we would like to discuss the physical 
significance of~$g$ with regard to the response 
characteristics of a cold atoms that are trapped 
in an optical billiard. We shall identify it as 
the dimensionless absorption coefficient, and we 
shall inquire the feasibility of witnessing the 
quantum $g_s$ suppression factor which is related   
to the connectivity of the induced Fermi-Golden-Rule (FGR) transitions.

\sect{LRT} 
In linear response theory one has to know the following 
information in order to calculate the EAR:
{\bf \ (i)} The temperature $T$ of the preparation;  
{\bf \ (ii)} The spectral fluctuations $\tilde{C}(\omega)$ of the system; 
{\bf \ (iii)} The spectral content $\tilde{S}(\omega)$ of the driving ;  
Let us elaborate on the latter. The RMS value 
of the vibrating wall velocity can be written as  
\be{0}
\dot{f} \mbox{\small [RMS]} \ \sim \ \omega_c A 
\ee
where $A$ is the amplitude of the wall movement. 
The power spectrum of $f(t)$ has a spectral 
support~$\omega_c$. To be specific let us assume that  
\be{91}
\tilde{S}(\omega) \ = \ \dot{f}^2 \, \frac{1}{2\omega_c}\exp\left(-\frac{|\omega|}{\omega_c}\right)
\ee
The wall vibrations induce diffusion in energy space.
Within LRT the diffusion coefficient is given by the Kubo formula, 
which in the following version can be regarded as an 
Einstein fluctuation-dissipation relation:  
\be{111}
D \ = \ \int_{0}^{\infty} \tilde{C}(\omega) \tilde{S}(\omega) d\omega
\ee
The EAR per particle for strongly chaotic dynamics, 
assuming that correlations between collisions can be neglected,  
is given by the {\em wall formula} \mycite{wall1,wall2,wall3}. Here we 
use the 2D version \mycite{frc}: 
\be{99}
\dot{E} 
\ =  \ \frac{D}{T} 
\ = \ \frac{1}{2T} \left[\frac{8}{3\pi} \frac{\mass^2 v_{\tbox{E}}^3}{L_x}\right] \dot{f}^2 
\ \equiv \ G_0\dot{f}^2
\ee
Regarding the ballistic period as the time unit, 
and $T$ as the energy unit, 
the dimensionless EAR is 
\be{9444}
\frac{\dot{E}}{T\Delta_{\tbox{L}}} 
\ = \ \frac{8}{3\pi^2} 
\left(\frac{\omega_c}{\Delta_{\tbox{L}}}\right)^2 
\left(\frac{A}{L}\right)^2  
\ee
In the quantum context the level spacing $\Delta_0$ sets the 
natural units for both energy and time measurements. 
Accordingly, we calculate the dimensionless quantity   
\be{0}
\frac{D}{\Delta_0^3} 
\ = \ \frac{8}{3\pi^2} 
\left(\frac{\Delta_{\tbox{L}}}{\Delta_0}\right)^3
\left(\frac{\omega_c}{\Delta_0}\right)^2
\left(\frac{A}{L}\right)^2  
\ee

\sect{FGR} 
The LRT formula \Eq{e111} can be obtained from a classical  
derivation, say using a kinetics Boltzmann picture, 
that does not assume applicability of the FGR picture. 
The same formula is obtained from FGR but with reservations 
that we illuminate in the next paragraph. It is therefore 
important to figure out the border between the quantum FGR regime 
and the classical Boltzmann regime.
The strict FGR condition states that the near-neighbor transitions 
between levels should have a rate ${w_0 < \omega_c}$.
Taking into account that the diffusion coefficient 
can be written as ${D \approx b_c \times w_0 \times \Delta_0^2}$, 
where $b_c = \omega_c/\Delta_0$, it follows that   
the strict FGR condition can be written as       
\be{9666}
\frac{D}{\Delta_0^3} \ \ < \ \ \left(\frac{\omega_c}{\Delta_0}\right)^{\mbox{power}}  
\ee
with power$=2$. But to witness FGR physics we can allow 
non-perturbative mixing on microscopic energy scales. 
The more careful analysis of \mycite{kbn} leads to the same 
condition but with power$=3$.

\sect{SLRT} 
It has been illuminated in a series of publications \mycite{kbr,slr,kbd} 
that in the FGR regime one should refer in general to semilinear response theory. 
SLRT applies to circumstances in which the environmental relaxation 
is weak compared with the $f(t)$-induced transitions. 
In such circumstances the connectivity of the transitions 
from level to level is important, and the diffusion coefficient 
is obtained via a resistor network calculation.
Let us give a more precise quantitative description of 
this latter statement. The absorption coefficient~$G$ 
is defined via \Eq{e1865}.
This is strictly analogous to Joule law: here the  
heating is due the vibration of the piston, while in 
the Joule-Drude problem it is due the oscillation 
of an electric field. 
The calculation of~$G$ can be done either within 
the framework of LRT using the Kubo formula (getting $G_{\tbox{LRT}}$), 
or within the framework of SLRT \mycite{kbr,slr,kbd}  
using a resistor-network calculation (getting $G_{\tbox{SLRT}}$). 
The correlations between collisions lead in the LRT case 
to a result that one can write as 
\be{0}
G_{\tbox{LRT}} \ \ = \ \ g_c G_0
\ee
where the expression for $G_0$ is implied by \Eq{e99}, 
and $g_c$ is defined as in \Eq{e300}.
Similarly it is convenient to write the outcome 
of the SLRT analysis as follows:
\be{0}
G_{\tbox{SLRT}} \ = \ g_s \, G_{\tbox{LRT}} \ = \ g_s \, g_c \, G_0 \ = \ g \, G_0 
\ee
where $g$ and $g_s$ are defined as in \Eq{e310} and \Eq{e29}.
If QCC considerations apply, then ${g_s\sim 1}$ 
with small $\hbar$ dependent corrections as in \Eq{e86}.

The results of SLRT differ from those of LRT 
if the perturbation matrix is either sparse or textured, 
which is the case if we have WQC circumstances.   
The LRT and SLRT numerical results for $g_c$ and for $g$ 
are displayed in \Fig{f34}.

\section{Experimental Manifestation of Quantum anomaly}

\label{s9}

With slight changes in notations which we find appropriate 
for the experimental context, we summarize again 
the main parameters of the problem:
\be{0}
\omega_{\tbox{L}} & \ \ = \ \ & \mbox{ballistic frequency} \\
\omega_{\tbox{R}} & \ \ = \ \ & \mbox{Lyapunov ergodization rate} \\
\omega_c  & \ \ = \ \ & \mbox{vibrations frequency span} \\
\omega_0  & \ \ = \ \ & \mbox{mean level spacing}
\ee
The length scales are the linear dimension $L$, 
the de~Broglie(thermal) wavelength $\lambda_{\tbox{E}}$ 
as determined by the temperature (calculated for ${E \sim T}$), 
and the radius of curvature of the walls~$R$.
The associated dimensionless parameters are: 
\be{0}
\hbarr &  = \  
\lambda_{\tbox{E}}/L &  \ 
= \mbox{dimensionless Planck} \\ 
u &   = \  
L/R  & \  
= \mbox{deformation parameter} \\
b &   = \  
u/h  & \  
= \mbox{dimensionless bandwidth} \\
a &  = \  
A/L  &  \ 
= \mbox{scaled vibration amplitude} 
\ee
Note that $u$ determines the ratio $\omega_{\tbox{R}}/\omega_{\tbox{L}}$, 
while $h$ determines the ratio $\omega_0/\omega_{\tbox{L}}$,  
hence $b=\omega_{\tbox{R}}/\omega_0$. 
Our interest is in the non-trivial possibility ${h < u \ll 1}$, 
else $\omega_{\tbox{R}}$ cannot be resolved.

\sect{The system} 
Following \mycite{nir1,nir3,nir4} we consider ${}^{85}Rb$ atoms (${\mass=1.4 \times 10^{-25} kg}$),
that are laser cooled to low temperature of {${T \approx 0.1\mu K}$}, 
such that the de~Broglie wavelength is {${\lambda_{\tbox{E}} = 1 \,\mu m}$}. 
The atoms are trapped in an optical billiard whose blue-detuned light walls 
confine the atoms by repulsive optical dipole potential. 
The motion of the atoms is limited to the billiard plane by a strong perpendicular 
optical standing wave. Assuming that the linear size of 
the billiard is {${L=10 \,\mu m}$},   
the dimensionless Planck is {$\hbarr=0.1$} leading 
to ${\omega_{\tbox{L}}/\omega_0=30}$. Note that 
\be{0}
\omega_{\tbox{L}} =& [2\pi] v_{\tbox{E}}/(2L) &= 220 \,\mbox{Hz} \\
\omega_0 =& [2\pi] \hbarr_{\tbox{plank}}^2/(\mass L^2)  &= 7.5 \,\mbox{Hz}
\ee
where the $[2\pi]$ should be omitted for Hz units.
Assuming $10\%$ deformation the dimensionless bandwidth 
can be tuned as {$b\sim 10$}.   

By modulating the laser intensity, 
one of the billiard walls can be noisily vibrated.
We assume that the driving is band-matched, 
i.e. $\omega_c\sim \omega_{\tbox{R}}$.
These are roughly the same parameters as in our analysis, 
for which we expect~{${g_s\sim 0.1}$}.

\sect{The SLRT anomaly} 
The common-wisdom expectation is that if QCC applies 
with regard to $\tilde{C}(\omega)$, then from \Eq{e111} 
we should get for the absorption coefficient roughly 
the same result classically and quantum mechanically. 
SLRT challenges this expectation. 
It applies to circumstances in which the environmental relaxation 
is weak compared with the $f(t)$-induced transitions. 
In such circumstances the connectivity of the transitions 
from level to level is important, and the LRT result should 
be multiplied by~$g_s$.    

In order to witness the SLRT anomaly, the driving amplitude~$A$ 
should be large enough so as to have a measurable heating effect, 
but small enough such that the FGR condition is not violated. 
Disregarding prefactors of order unity it follows from \Eq{e9444} and \Eq{e9666} 
that the requirements are 
\be{0}
a^2  & \ \ > \ \ & 10^{-3} \\
b^5 \times a^2 & \ \ < \ \ & b^3
\ee
The first condition is bases on the assumption that it 
is possible to hold the atoms for a duration of $\sim 1000$ bounces. 
Accordingly, there is a range where both conditions are satisfied, 
and there the SLRT anomaly should be observed, 
provided environmental relaxation effects can be neglected.

It is worth noting that our theory for $G$ is called SLRT because 
on the one hand ${\tilde{S}(\omega) \mapsto c \tilde{S}(\omega)}$ 
leads to ${G \mapsto c G}$, but on the other 
hand ${\tilde{S}(\omega) \mapsto \tilde{S}_1(\omega)+\tilde{S}_2(\omega)}$ 
does not lead to ${G \mapsto G_1 + G_2}$. 
This semi-linearity can be tested in an experiment 
in order to distinguish it from linear response.

\section{Ballistic versus diffusive scattering} 

\label{s10}
   
The EAR due to low frequency driving 
is determined by the couplings $|F_{nm}|^2$ 
between nearby levels. Let us see how conflicting 
expectations with respect to its dependence 
on~$u$ reconcile by the analysis that we have introduced.   
For a small deformation FOPT implies 
that the couplings are $\propto u^2$, and hence 
\be{0}
\dot{E}\propto u^2   \ \ \ \ \ \ \mbox{[FOPT expectation]}
\ee
As $u$ becomes larger the common expectation, 
based on Wigner theory, is to have Lorentzian mixing, 
leading to couplings $\propto 1/u^2$, and hence one expects 
\be{0}
\dot{E}\propto 1/u^2   \ \ \ \ \ \ \mbox{[Wigner expectation]}
\ee
In the formally equivalent problem of a conductance calculation this
``Joule law'' implies that the conductance is ${G \propto 1/u^2}$, 
where $u$ represents the strength of the disordered potential. 
For the purpose of derivation, 
instead of using the FGR or Wigner picture, 
one can use the equivalent Drude picture, 
where the Born mean free path is ${\ell \propto 1/u^2}$.    
On the other hand QCC considerations, based on \Eq{e2} and using \Eq{e42},  
imply that the couplings should be ${\propto 1/u}$, 
and hence one expects  
\be{0}
\dot{E}\propto 1/u   \ \ \ \ \ \ \mbox{[QCC expectation]}
\ee
We therefore encounter here 3 conflicting expectations for the 
dependence of the EAR on the deformation parameter. 
The analysis that we have presented resolves the conflict. 
Let us emphasize the main insights.

\sect{Ballistic scattering}
We have assumed a smooth deformation: the worst case 
was ${\alpha=1}$, but more generally we might have 
softer deformations with ${\alpha>1}$.
Consequently, the mixing is not uniform:
there are levels that are not mixed even if 
the perturbation is strong enough to mix some other levels.
This leads to an interesting co-existence 
of Semiclassical theory and FOPT.
Namely, we observe that the $\langle\langle |F_{nm}|^2 \rangle\rangle_a$
agrees with Semiclassics, 
while $\langle\langle |F_{nm}|^2 \rangle\rangle_s$ 
is given essentially by FOPT. 
The standard Wigner theory does not apply, 
and the EAR would is  $\propto u^2$ 
or $\propto 1/u$ depending on whether LRT or SLRT applies:
As the driving strength is increased we expect 
a crossover from LRT to SLRT.

\sect{Diffusive scattering}
If the deformation profile $D_u(y)$ is erratic on 
sub~$\lambda_{\tbox{E}}$ scale, then $U$ is somewhat  
similar to the white disorder that has been analyzed 
in Ref.\mycite{bld,kbd}. Under such circumstances 
all the matrix elements of $U_{nm}$ are {\em comparable}. 
Consequently, one would observe Lorentzian mixing $\propto u^2$. 
Therefore $\tilde{C}(\omega)$ would have a 
Lorentzian peak of width $\propto u^2$, which differs 
from the semiclassical peak $\propto u$. 
Furthermore, taking into account that 
the area under the central peak of $C(r)$ remains the same 
irrespective of~$u$, one deduces that    
\be{0}
\langle\langle |F_{nm}|^2 \rangle\rangle_{a/s} \propto \frac{1}{u^2}
\ \ \ \ \ \ \mbox{[Wigner mixing]}
\ee
and hence very different from both the FOPT prediction ${\propto u^2}$, 
and from the semiclassical expectation ${\propto u}$.   
In other words - for diffusive scattering, 
unlike ballistic scattering, QCC does not apply. 
If $U$ were like ``white disorder" the quantum dynamics 
would be characterized by the Born mean free path, 
which is very different from the classical mean free path.


\section{Summary}

It is important to realize that we are studying in this work 
a driven chaotic system, and not a driven integrable system. 
Remarkable examples for driven integrable systems are the 
kicked rotator \mycite{qkr1,qkr2,qkr3,qkr4} and the vibrating elliptical billiard \mycite{eb}.
In the absence of driving such systems are integrable, 
while in the presence of driving a {\em mixed phase space} emerges.      
This is not what we call here {\em weak chaos}. 
Rather our focus is on completely chaotic systems that have 
a very small Lyapunov exponent compared with the ballistic scale. 

Weakly chaotic systems do not fit the common RMT framework.  
The Hamiltonian matrix of such a driven system does not 
look like one that is taken from a Gaussian ensemble, 
but rather it is very sparse. 
One can characterize this sparsity by parameters~$s$ and~$g$
that reflect the percentage of large elements, and their connectivity, 
respectively. For~$g$ we have used a resistor network calculation that has 
direct relation to the semi-linear response characteristics of the system. 

We have highlighted that weakly chaotic systems possess 
a distinct WQC regime, much wider than originally expected, 
where semiclassics and Wigner-type mixing co-exist. 
Then we discussed the implications of this observation 
with regard to the theory of response.

The heating of particles in a box with vibrating walls 
is a prototype problem for exploring the limitations 
of linear response theory and the quantum-to-classical 
correspondence principle. In the experimental arena 
this topic arises in the theory of {\em nuclear friction} \mycite{wall1,wall2,wall3}, 
and in the studies of cold atoms 
that are trapped in {\em optical billiards} \mycite{nir1,nir2,nir3,nir4}.   
Mathematically it is related to the analysis
of mesoscopic conductance of ballistic rings \mycite{bld}.   
In typical circumstances the classical analysis predicts 
an absorption coefficient that is determined by the Kubo formula 
\mycite{ott1,ott2,ott3,wilk,WA,robbins,jar1,jar2,frc}, 
leading to the ``Wall formula" in the nuclear context, 
or to the analogous ``Drude formula" in the mesoscopic context.     
The question arises \mycite{wilk,WA,robbins,frc,crs,rsp,krav1,krav2,kbr,slr,kbw} 
are there circumstance in which the quantum theory leads to a novel 
result that does not resemble the semiclassical prediction.

The low frequency driving that we assume is stochastic, 
rather than periodic. This looks to us realistic, 
reflecting the physics of cold atoms that are trapped in optical billiards with vibrating walls. 
It is also theoretically convenient, because we can use the Fermi-Golden-Rule picture. 
If one is interested in periodic driving of strictly isolated system, 
then there are additional important questions  
with regard to dynamical localization \mycite{qkr1,qkr2,qkr3,qkr4,lc}, 
that can be handled e.g. within the framework of the Floquet theory approach.

We predict that the EAR of a weakly chaotic system
in the WQC regime would exhibit an SLRT anomaly: 
An LRT to SLRT crossover is expected as the intensity of the driving is increased;  
the linearity with respect to the intensity of the source is maintained but 
with a different (smaller) coefficient;   
while the linearity with respect to the addition of independent sources is lost. 
This anomaly reflects that the absorption process in the mesoscopic 
regime might resemble a percolation process due to the sparsity 
of the perturbation matrix. In systems with diffusive scattering, 
that are in the focus of standard condense matter textbooks, 
such an effect could not arise.


\sect{Acknowledgements}
We thank Nir Davidson (Weizmann) for a crucial 
discussion regarding the experimental details.  
This research has been supported by the US-Israel Binational Science Foundation (BSF).

\appendix

\section{The resistor-newtwork average}

\label{a0}

We use the notation $\langle\langle \bm{X} \rangle\rangle$
in order to indicate the {\em average} value of its {\em in-band} 
elements. First we would like to define the standard 
{\em algebraic} average. It is essential to introduce a {\em weight} 
function that defines the band of interest. In the physical 
context this function reflects the spectral content of the 
driving sources. In practice we use rectangular 
or exponential weight function, say 
\be{0}
F(r) \ \ = \ \ \frac{1}{2b_c}\eexp{-|n-m|/b_c} 
\ee 
which corresponds to \Eq{e91}.
For characterization purpose we assume 
a band-matching weight function, 
meaning that $b_c$ is chosen as the natural bandwidth 
of the matrix, corresponding to $\Delta_{\tbox{R}}$.
The algebraic average is defined in the standard way:
\be{0}
\langle\langle \bm{X} \rangle\rangle_a \ \ = \ \ 
\frac{1}{N} \sum_{n,m} F(n-m) \ X_{nm}
\ee 
where $N$ is the size of the matrix, 
which is assumed to be very large.  
The algebraic average is a {\em linear} operation, 
meaning that  
\be{0}
\langle\langle \lambda \bm{X} \rangle\rangle &=& \lambda \langle\langle \bm{X} \rangle\rangle
\\
\langle\langle \bm{X} + \bm{Y} \rangle\rangle &=& 
\langle\langle \bm{X} \rangle\rangle + \langle\langle \bm{Y} \rangle\rangle 
\ee

There are different type of ``averages" in the literature, 
such as the {\em harmonic} average,  {\em geometric} average, 
and we can also include the {\em median} in the same list.
All these ``averages" are {\em semi-linear} operations because
only the  $\langle\langle \lambda \bm{X} \rangle\rangle = \lambda \langle\langle \bm{X} \rangle\rangle$ 
property is satisfied for them.   
Irrespective of the semi-linearity issue {\em any} type of 
average should satisfy the following requirement:  
if all the elements equal to the same number, 
then also the average should equal the same number.
 
In this paper we highlight a new type of average 
that we call a {\em resistor-newtwork} average. 
The defining prescription for its calculation is simple: 
given $X_{nm}$  we associate with it a resistor network $G_{nm}$
via \Eq{e28}, and define $\langle\langle \bm{X} \rangle\rangle_s$
as its inverse resistivity.

There are a few cases where an analytical expression 
is available for the inverse resistivity $G$ of a network $G_{nm}$. 
If only near neighbor nodes are connected, 
allowing ${G_{n,n+1}=g_n}$ to be 
different from each other, 
then ``addition in series" implies 
that the inverse resistivity calculated 
for a chain of length~$N$ is  
\be{0}
G \ \ = \ \ \left[\frac{1}{N}\sum_{n=1}^{N} \frac{1}{g_n}\right]^{-1}
\ee
If $G_{nm}=g_{n-m}$ is a function 
of the distance between the nodes $n$ and $m$ 
then it is a nice exercise to prove 
that ``addition in parallel" implies 
\be{0}
G \ \ = \ \ \sum_{r=1}^{\infty} r^2 g_r
\ee
Note that in the latter case the resistor network 
average {\em coincides} with the algebraic 
average. In order to have a different 
result the diagonals of the matrix should 
be non-uniform, which is the case for sparse 
or textured matrices. 
 
In general an analytical formula for~$G$
is not available, and we have to apply 
a numerical procedure. For this purpose 
we imagine that each node~$n$ is connected 
to a current source $I_n$. The Kirchhoff    
equations for the voltages are    
\be{0}
\sum_m G_{mn} (V_n-V_m)  \  = \ I_n
\ee
This set of equation can be written in a matrix form:
\be{0}
\bm{G} \bm{V}  \ = \ \bm{I}
\ee
where the so-called discrete Laplacian matrix of the network is defined as 
\be{0}
\bm{G}_{nm} =  \left[\sum_{n'}  G_{n'n}\right]\delta_{n,m} - G_{nm}
\ee
This matrix has an eigenvalue zero which is associated  
with a uniform voltage eigenvector. Therefore, it has 
a pseudo-inverse rather than an inverse, and the Kirchhoff 
equation has a solution if and only if ${\sum_n I_n=0}$.      
In order to find the resistance between nodes ${n_{\tbox{in}}=0}$ 
and ${n_{\tbox{our}}=N}$, 
we set ${I_0=1}$ and ${I_N=-1}$ and ${I_n=0}$ otherwise,  
and solve for $V_0$ and $V_N$. 
The inverse resistivity is ${G=[(V_0-V_N)/N]^{-1}}$.

\section{Intensity of fluctuations - optional derivations}

\label{a1}

In this appendix we clarify the low frequency 
behavior of $\tilde{C}(\omega)$ using two optional 
approaches. We assume that $\gamma\sim\gamma_0$ 
is roughly a constant, so there is a well defined correlation time 
\be{0}
t_{\tbox{R}} = \frac{1}{\gamma} 
\ee

\sect{Time domain approach}
Observe that  
\be{0}
\int_{-0}^{t} {F_{\theta}(t)} dt = 
\begin{cases}
q_{\theta}, & t \ll t_{\tbox{L}} \\
(q_{\theta}/\tau_{\theta})t, &   t_{\tbox{L}} \ll t \ll t_{\tbox{R}} \\
\langle q_{\theta}/\tau_{\theta}\rangle t,  & t \gg t_{\tbox{R}}
\end{cases}
\ee
From \Eq{e34} it follows that $\langle F(0)F(t) \rangle$ looks as follows:  
at ${t=0}$ it contains a self-correlation delta peak;
within ${t \ll t_{\tbox{R}}}$ it is the theta averaged comb of delta peaks due to bouncing; 
For ${t \gg t_{\tbox{R}}}$ it flattens and reflects the squared average value of $F$. 
Accordingly, the short time average and the long time average values 
of $\langle F(0)F(t) \rangle$ are $c_0$ and $c_{\infty}$ of \Eq{e30} and \Eq{e31}.
Consequently, the ``area" under the correlation function is  
\be{0}
\tilde{C}(\omega{=}0) &\approx& (c_0{-}c_{\infty}) \times 2t_{\tbox{R}} 
\ee
in agreement with \Eq{e42}.

\sect{Number variance approach}
It is instructive to deduce $\tilde C(\omega=0)$ 
using over-simplified derivation 
via the number variance approach,  
as in the analysis of spectral rigidity \mycite{berry}.
This over-simplified approach treats 
the spikes as having equal size (below $q=1$). 
The variance in the number of collisions during
the time~$t$ is given by the expression   
\be{0}
\mbox{Var}(N(t)) = 
\frac{2}{\pi^2} \int_0^{\infty} \frac{\tilde C(\omega)}{\omega^2}
\sin^2(\pi \omega t) d\omega
\ee
Consequently, 
\be{0}
\tilde C(\omega=0) = \frac{\mbox{Var}(N(t))}{t} =\mbox{diffusion in counting} 
\ee
Assuming that the step in this 
random walk process is of duration $t_{\tbox{R}}$,  
the diffusion coefficient is  
\be{0}
\tilde C(\omega=0) 
=  \frac{1}{t_{\tbox{R}}} \mbox{Var}\left( \frac{t_{\tbox{R}}}{\tau} \right)
= \frac{1}{\gamma} \mbox{Var}\left( \frac{1}{\tau} \right)
\ee
Which leads upon restoration of $q$ to \Eq{e42}.

\section{The $F_{nm}$ matrix for zero deformation}

\label{a2}

Here we calculate the large scale sparsity~$p_0$, 
and the average value of $|F_{nm}|^2$, 
in the case of a rectangular box. 
It is tempting to identify $p_0$ as "$s$" , 
but in fact the latter is ill defined because 
it refers to the sparsity of the in-band elements, 
while for $u=0$ the bandwidth $\Delta_{\tbox{R}}$ is {\em zero}.  

We consider the matrix elements that reside 
inside an energy window of width $\delta E$.
The levels ${(n_x,n_y)}$ within this window 
belong to the energy shell ${E<E_{n_x,n_y}<E+\delta E}$. 
We define the ``radius" of this shell 
as  ${k_{\tbox{E}} = (2\mass E)^{1/2}}$. 
For a given $n_y$ section the width 
of the shell is denoted as $\delta n_x$, 
and in wavenumber units it is given by the expression  
\be{0}
\delta k_x = \delta \sqrt{k_{\tbox{E}}^2-k_y^2} 
\approx \frac{k_{\tbox{E}} \delta k_{\tbox{E}}}{\sqrt{k_{\tbox{E}}^2-k_y^2}}
\ee
The total number of levels within this window 
can be calculated in a complicated way as 
\be{0}
\mathcal{N} 
&=& 
\int_0^{k_{\tbox{E}}} 
\delta n_x dn_y 
\\
&=&  
\frac{L_xL_y}{\pi^2}
\int_0^{k_{\tbox{E}}} 
\frac{\mass\delta E}{\sqrt{k_{\tbox{E}}^2-k_y^2}} dk_y
\ = \ 
\frac{\delta E}{\Delta_0}
\ee
Similarly we can calculate the number of coupled levels, 
and hence the large scale sparsity:
\be{0}
p_0
&=&
\frac{1}{\mathcal{N}^2} 
\int_0^{k_{\tbox{E}}} 
\delta n_x^2 dn_y 
\\
&=&  \nonumber
\frac{4}{\pi L_y}
\int_0^{k_{\tbox{E}}} 
\frac{1}{k_{\tbox{E}}^2-k_y^2} dk_y
\ = \ 
\frac{2}{\pi k_{\tbox{E}}L_y}\ln\left[\frac{2 k_{\tbox{E}}}{dk}\right]
\ee

The FOPT perturbed matrix is sparse and textured. 
Its non-zero elements are of size  ${k_x^2 /\mass L_x}$
as implied by \Eq{e47}. The algebraic average of the elements 
is given by 
\be{0}
&&
\langle\langle |F_{nm}|^2 \rangle\rangle_{\infty} 
 =
\frac{1}{\mathcal{N}^2} 
\int_0^{k_{\tbox{E}}} 
\delta n_x^2 dn_y 
\left[\frac{k_x^2} {\mass L_x}\right]^2 
\\ \nonumber 
&& \ \ = 
\frac{4}{\pi L_y}
\int_0^{k_{\tbox{E}}}
\frac{1}{k_{\tbox{E}}^2-k_y^2} dk_y
\left[\frac{k^2 - k_y^2} {\mass L_x}\right]^2 
=
\frac{8}{3\pi}\frac{k_{\tbox{E}}^3}{\mass^2 L_x^2L_y}
\ee
which is the same result \Eq{e19}
as in the semiclassical estimate.



\clearpage

\begin{figure}[h!]
\includegraphics[clip,width=\hsize]{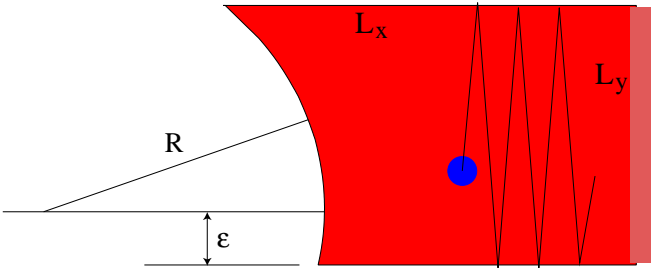}

\caption{(Color online)
Sketch of the billiard system of \Eq{e1}. 
The unperturbed billiard is a rectangle of size $L_x \times L_y$.  
The deformation $U$, due to the radius of curvature $R$ of the left wall,  
is characterized by the parameter ${u=L_y/R}$. 
In order to break the mirror symmetry the center of the curved wall 
is shifted upwards a vertical distance ${\varepsilon}$.   
The time dependent perturbation is due to the displacement $f(t)$ of the right wall. 
}

\mylabel{f11}
\end{figure}

\begin{figure}[h!]
\includegraphics[clip,width=0.8\hsize]{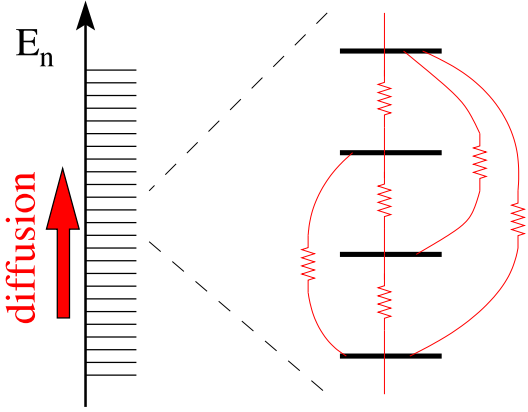}

\caption{(Color online)
The driving induces transitions between levels $E_n$ of a closed system, 
leading to diffusion in energy space, and hence an associated heating. 
The diffusion coefficient $D_{\tbox{E}}$ can be calculated using a resistor network analogy. 
Connected sequences of transitions are essential in order 
to have a non-vanishing result, as in the theory of percolation.
}

\mylabel{f10}
\end{figure}

\begin{figure}[h!]
\includegraphics[clip,width=0.85\hsize]{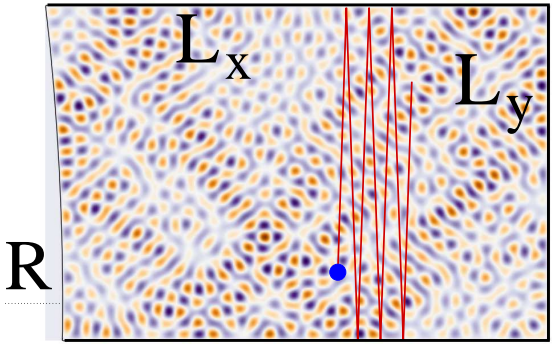}

\caption{(Color online)
An image of the eigenstate $E_n{\simeq}13618$ 
for the billiard of \Fig{f11} with $L_x{=}1.5$, 
and $L_y{=}1.0$, and ${R{=}8}$, and ${\varepsilon=0.1}$.
In the numerics the units are chosen 
such that $\hbar_{\tbox{Planck}}{=}1$ 
and the mass is $\mass{=}1/2$. 
}

\mylabel{f12}
\end{figure}

\newpage


\begin{figure}[H]
\includegraphics[clip,width=0.8\hsize]{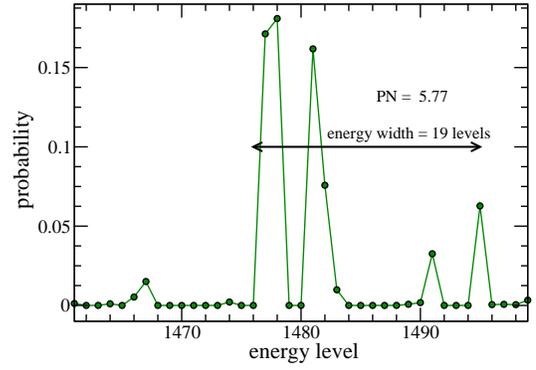}

\caption{(Color online)
Using a truncated matrix representation of 
the deformed billiard ${\mathcal{H}_0+U}$ 
in the unperturbed basis ${\bar{\bm{n}}=(n_x,n_y)}$ of the 
non-deformed rectangular billiard ${\mathcal{H}_0}$, 
we find a representative eigenstate~$n_0$ of the former 
and plot $|\langle E_{\bar{\bm{n}}} | E_{n_0} \rangle|^2$
versus the running index $\bar{\bm{n}}$ of the ordered energies.  
The participation number (PN) of an eigenstate in this basis 
reflects how many energy levels were mixed due to the deformation~$u$.
Having PN${>}1$ indicates that our data is beyond the FOPT regime.
For the displayed eigenstate $\mbox{PN}\approx 5.77$,   
while its energy width is $19$~levels. 
The billiard parameters are $R=8$ and $1/\hbarr \approx 27.15$. }

\mylabel{f13}
\end{figure}


\begin{figure}[H]
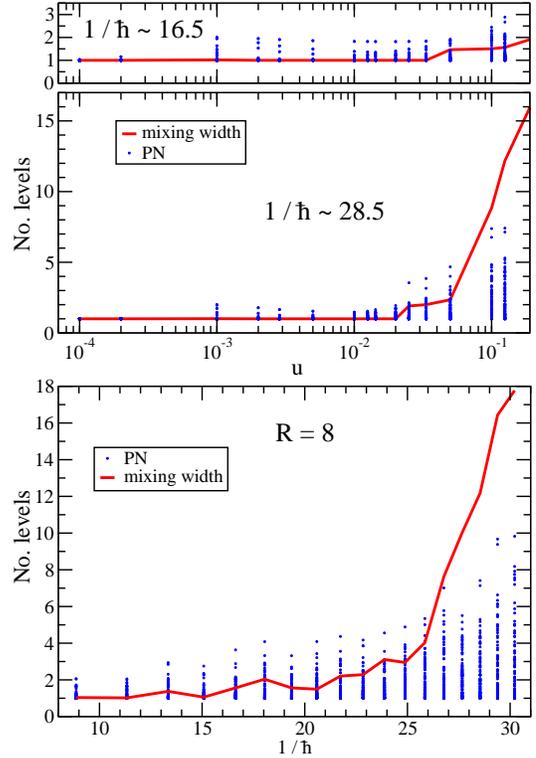

\includegraphics[clip,width=0.8\hsize]{cqc_PNwidth_vs_u}

\includegraphics[clip,width=0.8\hsize]{cqc_PNwidth_vs_hbar}

\caption{(Color online)
The participation number (PN) of the eigenstates 
as a function of $u$ (panel~a) and $\hbarr$ (panel~b). 
The parameter $\hbarr$ characterizes an energy window 
that contains $\sim100$ eigenstates.  
The method of calculation is as explained in \Fig{f13}.
The average energy width in units of mean level spacing 
is represented by red solid line. 
Having average width larger than the average PN is an indication for sparsity.}

\mylabel{f14}
\end{figure}

\newpage

\begin{figure}[h!]
\includegraphics[clip,width=\hsize]{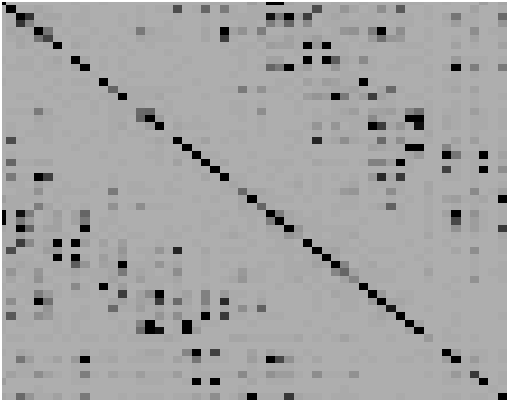}

\caption{(Color online)
Image of the perturbation matrix $\bm{X}=\{|F_{nm}|^2\}$ 
for the billiard of \Fig{f12}.
within the energy window ${3500< E_n < 4000}$. 
This matrix is {\em sparse}. More generally 
it might have some {\em texture}. The latter 
term applies if the arrangement of the large 
elements is characterized by some pattern. 
}

\mylabel{f31}
\end{figure}

\begin{figure}[h!]
\centering
\includegraphics[clip,width=\hsize]{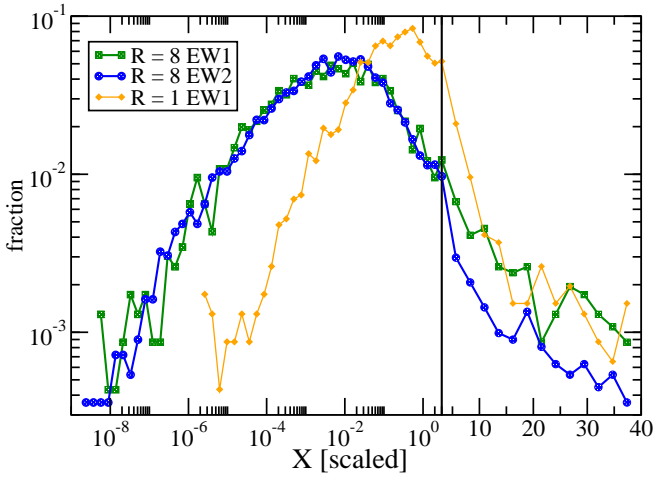}

\caption{(Color online)
Histogram of the values of the elements $X_{nm}$  that reside in 
the central band of the matrix. The analysis is done 
for the billiard of \Fig{f12} where~${R=8}$. 
The statistics includes all the elements in the 
energy windows ${100<E<4000}$ (EW1), and ${10000<E<14000}$ (EW2). 
For sake of comparison we display results also for~${R=1}$.  
}

\mylabel{f32}
\end{figure}

\newpage

\begin{figure}
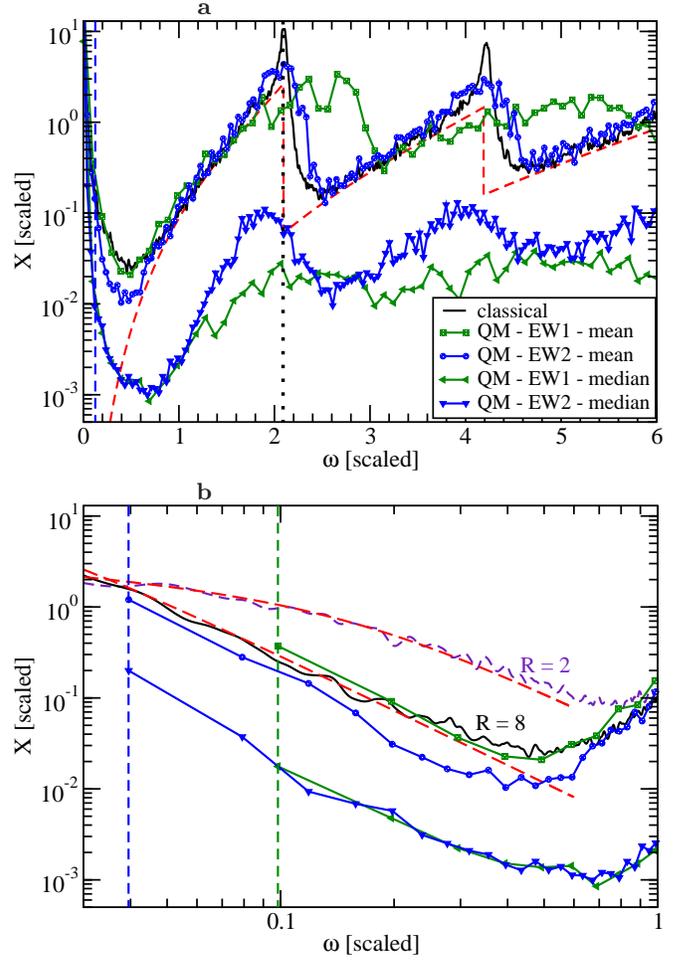

{\bf a} \hspace*{0.4\hsize} \\ 
\includegraphics[clip,width=\hsize]{cqc_F_alg_med_vs_omega_deform} \\
{\bf b}  \hspace*{0.4\hsize}  \\ 
\includegraphics[clip,width=\hsize]{cqc_F_alg_med_vs_omega_deform_zoom}

\caption{(Color online)
The band profile of the perturbation matrix 
for the billiard of \Fig{f12} where~${R=8}$. 
(a)~ 
The algebraic average and median along the diagonals 
of the $X_{nm}$ matrix versus ${\omega\equiv (E_n{-}E_m)}$. 
The vertical axis is normalized with respect to $C_{\infty}$, 
while the horizontal axis is $\omega/v_{\tbox{E}}$. 
The classical power spectrum is presented to demonstrate 
the applicability of the semiclassical relation \Eq{e2}.
The red line is the analytical expression 
that applies to zero deformation.
The dotted vertical line is the frequency $1/t_{\tbox{L}}$ and the dashed one is $1/t_{\tbox{R}}$.
(b)~ 
Zoom of the $\omega\ll 1/t_{\tbox{L}}$ region.
For sake of comparison we display results also for~${R=2}$.  
The vertical lines indicate the mean level spacing.
The dashed red curves are based on Eq.(\ref{e480}).
}

\mylabel{f33}

\end{figure}

\begin{figure}
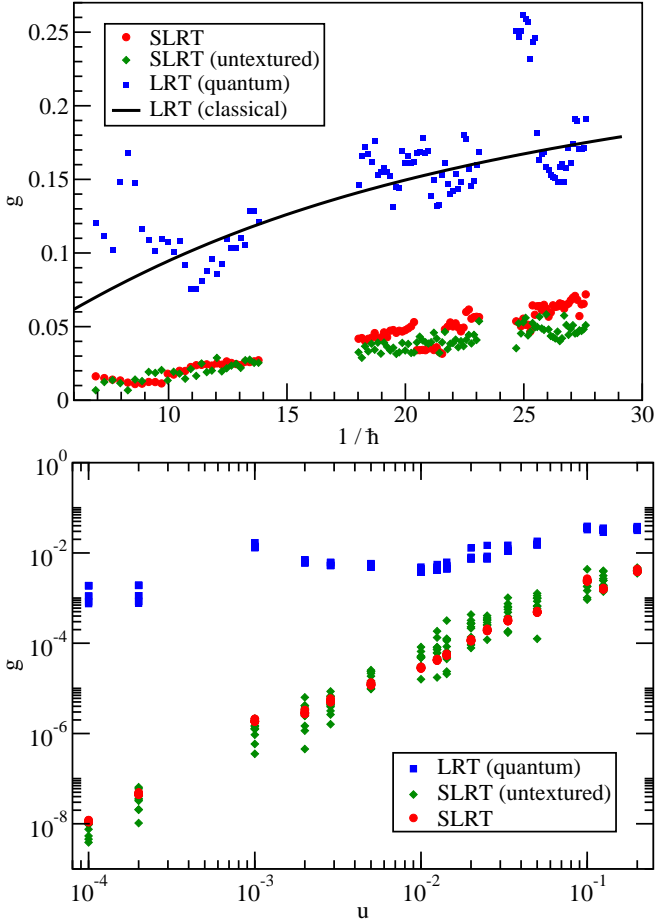

\centering
\includegraphics[clip,width=\hsize]{cqb_g_vs_h}

\includegraphics[clip,width=\hsize]{cqc_g_vs_u}

\caption{(Color online)
{\bf SLRT vs LRT.}
The scaled absorption coefficient $g_c$ (LRT) and $g=g_sg_c$ (SLRT) 
versus the dimensionless $1/\hbarr$ (upper panel),
and versus the dimensionless deformation parameter ${u=L/R}$ (lower panel).
Note that ${g=1}$ is the prediction of the ``Wall formula", 
while the line is based on the {\em classical} analysis.
In the upper panel the analysis has been done for the billiard of \Fig{f11}. 
The calculation of each point has been carried out on a $100 \times 100$ 
sub-matrix of $\bm{X}$  centered around the $\hbarr$ implied energy~$E$. 
The ``untextured'' data points are calculated for an artificial  
random matrices with the same bandprofile and sparsity (but no texture).  
The complementary lower panel is oriented to show the small~$u$ 
dependence. The analysis is based on a truncated matrix 
representation of ${\mathcal{H}_0+U}$,  
within an energy window that corresponds to ${1/\hbarr\sim9}$. 
Due to the truncation there is some quantitative inaccuracy 
with regard to the larger~$g$ values.   
}

\mylabel{f34}
\end{figure}


\begin{figure}[h!]
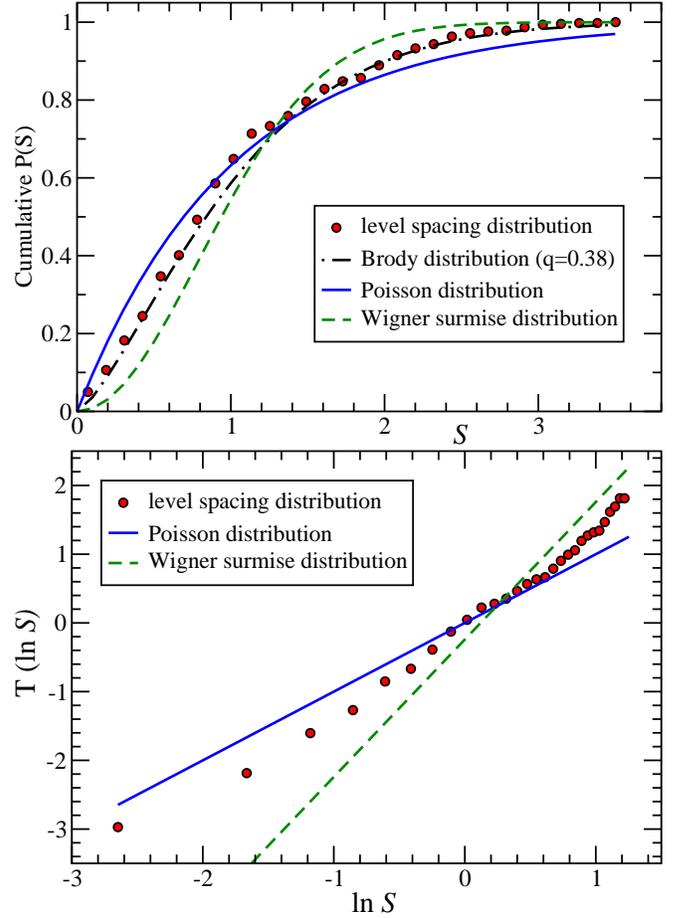

\includegraphics[clip,width=\hsize]{cqc_cumhist_Sn}

\includegraphics[clip,width=\hsize]{cqc_T_vs_lnSn}

\caption{(Color online)
(a) Cumulative histogram for the level spacing distribution~$P(S)$
with fitting to Brody distribution ($q{=}0.38$), 
and contrasted with Poisson distribution ($q{=}0$) 
and the Wigner surmise ($q{=}1$).
(b) The Brody parameter is determined via the slope 
of $T(x)$ as explained in the text.   
}

\mylabel{f21}
\end{figure}


\begin{figure}[h!]
\includegraphics[clip,width=\hsize]{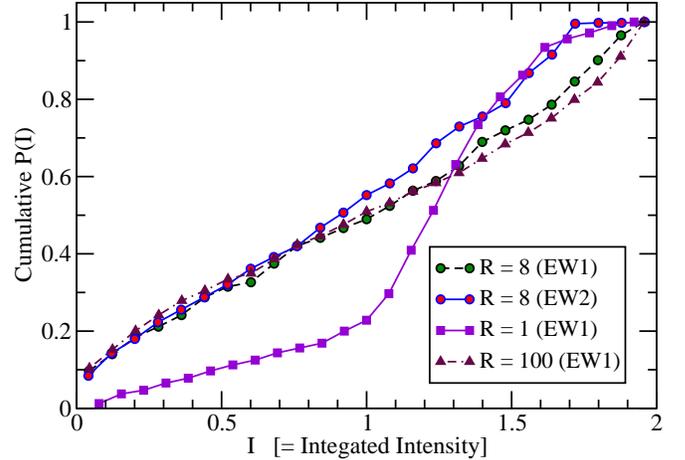}

\caption{(Color online)
Wavefunction intensity statistics:
The cumulative distribution of the integrated 
intensity $I_n$ of \Eq{e100} is presented.
As $R$ becomes larger the distribution 
further deviates from Gaussian statistics. 
The statistics includes all the eigenfunctions 
in the energy windows ${100<E<4000}$ (EW1), 
and ${10000<E<14000}$ (EW2). 
}

\mylabel{f22}
\end{figure}

\clearpage


\begin{thebibliography}{0}




\bibitem{mario1} 
M. Feingold, A. Peres, 
{\em Phys. Rev. A} {\bf 34} 591, (1986).

\bibitem{mario2} 
M. Feingold, D. Leitner, M. Wilkinson, 
{\em Phys. Rev. Lett.} {\bf 66}, 986 (1991). 
%
\bibitem{mario3} 
M. Wilkinson, M. Feingold, D. Leitner, 
{\em J. Phys. A} {\bf 24}, 175-182 (1991). 

\bibitem{mario4} 
M. Feingold, A. Gioletta, F.M. Izrailev, L. Molinari, 
{\em Phys. Rev. Lett.} {\bf 70}, 2936–2939 (1993).

\bibitem{prosen1} 
T. Prosen and M. Robnik, J. Phys. A 26 L319 (1993)

\bibitem{prosen2} 
T. Prosen, Ann. Phys. (N.Y.) 235, 115 (1994)




\bibitem{wall1}
D.H.E. Gross,
{\em Nucl. Phys. A} {\bf 240}, 472 (1975).

\bibitem{wall2}
J. Blocki, Y. Boneh, J.R. Nix, J. Randrup, M. Robel, A.J. Sierk, W.J. Swiatecki, 
{\em Ann. Phys.} {\bf 113}, 330 (1978).

\bibitem{wall3}
S.E. Koonin, R.L. Hatch, J. Randrup, 
{\em Nuc. Phys. A} {\bf 283}, 87 (1977). 




\bibitem{nir1} 
N. Friedman, A. Kaplan, D. Carasso, N. Davidson, 
{\em Phys. Rev. Lett.} {\bf 86}, 1518 (2001).

\bibitem{nir2}
A. Kaplan, M. Andersen, N. Friedman, N. Davidson, in 
{\em Chaotic Dynamics and Transport in Classical and Quantum Systems},
Editors: P. Collet, M. Courbage, S. Metens, A. Neishtast, G. Zaslavsky,  
NATO science series II, vol.182, p.239 (Springer 2004).

\bibitem{nir3}
A. Kaplan, N. Friedman, M. F. Andersen, and N. Davidson, Phys. Rev. Lett. 87, 274101 (2001).

\bibitem{nir4}
M. Andersen, A. Kaplan, T. Grunzweig and N. Davidson, Phys. Rev. Lett. 97, 104102 (2006).

\bibitem{kbw}
A. Stotland, D. Cohen, N. Davidson, 
{\em Europhys. Lett.} {\bf 86}, 10004 (2009).





\bibitem{bunimovich}
L. A. Bunimovich, and Ya. G. Sinai, Comm. Math. Phys. 78, 479-497, 1981

\bibitem{young}
L.-S. Young, Ann. of Math. 147(3), 585-650, 1998

\bibitem{bouncing1}
B.J. Alder, T.E. Wainwright, 
{\em Phys. Rev. A} {\bf 1}, 18 (1970).

\bibitem{bouncing2}
F. Vivaldi, G. Casati, I. Guarneri, 
{\em Phys. Rev. Lett.} {\bf 51}, 727 (1983).



\bibitem{triang}
G. Casati and T. Prosen, Phys. Rev. Lett. {\bf 85}, 4261 (2000)
M. Degli Esposti, S. O’Keefe and B. Winn, Nonlinearity {\bf 18}, 1073 (2005).

\bibitem{spectral}
E.B. Bogomolny, U. Gerland, C. Schmit, 
Phys. Rev. E {\bf 59}, R1315 (1999).

\bibitem{Backer}
A. Backer, R. Schubert, P. Stifter,
J. Phys. A {\bf 30} 6783 (1997).

\bibitem{wqc1}
F. Borgonovi, G. Casati and B. Li,   
Phys. Rev. Lett. {\bf 77}, 4744 (1996). 

\bibitem{wqc2}
K.M. Frahm and D.L. Shepelyansky,
Phys. Rev. Lett. {\bf 78}, 1440 (1997).

\bibitem{cqb}
A. Stotland, L.M. Pecora and D. Cohen, Europhys. Lett. {\bf 92}, 20009 (2010). 




\bibitem{sparse1} 
E.J. Austin, M. Wilkinson, 
{\em Europhys. Lett.} {\bf 20}, 589 (1992). 

\bibitem{sparse2} 
T. Prosen, M. Robnik, 
{\em J. Phys. A} {\bf 26}, 1105 (1993).

\bibitem{sparse3} 
Y. Alhassid, R.D. Levine, 
{\em Phys. Rev. Lett.} {\bf 57}, 2879 (1986).

\bibitem{sparse4}
Y.V. Fyodorov, O.A. Chubykalo, F.M. Izrailev, G. Casati, 
{\em Phys. Rev. Lett.} {\bf 76}, 1603 (1996).




\bibitem{kbr}
D. Cohen, T. Kottos, H. Schanz, 
{\em J. Phys. A} {\bf 39}, 11755 (2006). 

\bibitem{slr}
M. Wilkinson, B. Mehlig, D. Cohen, 
{\em Europhys. Lett.} {\bf 75}, 709 (2006).

\bibitem{kbd}
A. Stotland, T. Kottos, D. Cohen, 
{\em Phys. Rev. B} {\bf 81}, 115464 (2010), 
and further references therein.




\bibitem{boundary}
R. Ram-Mohan,  
{\em Finite Element and Boundary Element Applications in Quantum Mechanics}  
(Oxford University Press, Oxford, UK, 2002).







\bibitem{frc}
D. Cohen, 
{\em Annals of Physics} {\bf 283}, 175 (2000).

\bibitem{dil} 
A. Barnett, D. Cohen, E.J. Heller, 
{\em Phys. Rev. Lett.} {\bf 85}, 1412 (2000);  

\bibitem{wlf}
A. Barnett, D. Cohen, E.J. Heller, 
{\em J. Phys. A} {\bf 34}, 413 (2001). 



\bibitem{kamenev} 
For a review see 
``(Almost) everything you always wanted to know about 
the conductance of mesoscopic systems" 
by A. Kamenev and Y. Gefen, Int. J. Mod. Phys. {\bf B9}, 751 (1995).



\bibitem{prm}
D. Cohen, A. Barnett, E.J. Heller, 
{\em Phys. Rev. E} {\bf 63}, 46207 (2001).



\bibitem{bld} 
A. Stotland, R. Budoyo, T. Peer, T. Kottos, D. Cohen, 
{\em J. Phys. A} (FTC) {\bf 41}, 262001 (2008). 




\bibitem{qkr1}
B.V.Chirikov, Phys. Rep. {\bf 52}, 263 (1979).

\bibitem{qkr2}
S. Fishman, D.R. Grempel and R.E. Prange, Phys. Rev. Lett. 49, 509 (1982).

\bibitem{qkr3}
S. Fishman in "Quantum Chaos",
{\em Proceedings of the International School
of Physics "Enrico Fermi", Course CXIX},
Ed. G. Casati, I. Guarneri and U. Smilansky
(North Holland 1991).

\bibitem{qkr4}
M. Raizen in "New directions in quantum chaos",
{\em Proceedings of the International School
of Physics "Enrico Fermi", Course CXLIII},
Edited by G. Casati, I. Guarneri and U. Smilansky
(IOS Press, Amsterdam 2000).

\bibitem{eb}
F. Lenz, F.K. Diakonos, P. Schmelcher,
Phys. Rev. Lett. {\bf 100}, 014103 (2008); 
Europhys. Lett. {\bf 79}, 2002 (2007).



\bibitem{kbn}
I. Sela, J. Aisenberg, T. Kottos and D. Cohen, J. Phys. A (FTC) {\bf 43}, 332001 (2010).  



\bibitem{ott1}
E. Ott, 
{\em Phys. Rev. Lett.} {\bf 42}, 1628 (1979). 

\bibitem{ott2}
R. Brown, E. Ott, C. Grebogi, 
{\em Phys. Rev. Lett.} {\bf 59}, 1173 (1987).

\bibitem{ott3}
R. Brown, E. Ott, C. Grebogi, 
{\em J. Stat. Phys.} {\bf 49}, 511 (1987).

\bibitem{jar1}
C. Jarzynski, 
{\em Phys. Rev.} {\bf E 48}, 4340 (1993).

\bibitem{jar2}
C. Jarzynski, 
{\em Phys. Rev. Lett.} {\bf 74}, 2937 (1995).

\bibitem{wilk}
M. Wilkinson, 
{\em J. Phys. A} {\bf 21}, 4021 (1988).

\bibitem{WA}
M. Wilkinson, E.J. Austin, 
{\em J. Phys. A} {\bf 28}, 2277 (1995). 

\bibitem{robbins}
J.M. Robbins, M.V. Berry, 
{\em J. Phys. A} {\bf 25} L961 (1992). 



\bibitem{crs}
D. Cohen, 
{\em Phys. Rev. Lett.} {\bf 82}, 4951 (1999). 

\bibitem{rsp}
D. Cohen, T. Kottos, 
{\em Phys. Rev. Lett.} {\bf 85}, 4839 (2000).

\bibitem{krav1}
D.M. Basko, M.A. Skvortsov, V.E. Kravtsov, 
{\em Phys. Rev. Lett.} {\bf 90}, 096801 (2003).

\bibitem{krav2}
A. Silva, V.E. Kravtsov, 
{\em Phys. Rev. B} {\bf 76}, 165303 (2007).

\bibitem{lc}
T. Prosen, D.L. Shepelyansky, Eur. Phys. J. B {\bf 46}, 515 (2005).  



\bibitem{berry}
M.V. Berry, Nonlinearity 1, 399 (1988)


\end{thebibliography}
\end{document}